\documentclass[aps,prd,reprint,superscriptaddress,preprintnumbers,nofootinbib]{revtex4-1}
\usepackage{amsmath}
\usepackage{amsfonts} 
\usepackage{amssymb}
\usepackage{graphicx}
\usepackage{hyperref}
\usepackage{tensor}
\usepackage{siunitx}
\usepackage{xcolor}

\newcommand{\be}{\begin{equation}}
\newcommand{\ee}{\end{equation}}

\newcommand{\beq}{\begin{equation}}  \newcommand{\eeq}{\end{equation}}
\newcommand{\bal}{\begin{aligned}}   \newcommand{\eal}{\end{aligned}}
\newcommand{\bea}{\begin{eqnarray}}  \newcommand{\eea}{\end{eqnarray}}

\begin{document}

\title{\Large AdS and the Swampland}
\preprint{MPP-2019-116, LMU-ASC 24/19}
\preprint{}

             \author{\large Dieter L\"ust}
\affiliation{Arnold-Sommerfeld-Center for Theoretical Physics, Ludwig-Maximilians-Universit\"at, 80333 M\"unchen, Germany}
\affiliation{Max-Planck-Institut f\"ur Physik (Werner-Heisenberg-Institut),
             F\"ohringer Ring 6,
             80805, M\"unchen, Germany}
\author{\large Eran Palti}
\affiliation{Max-Planck-Institut f\"ur Physik (Werner-Heisenberg-Institut),
             F\"ohringer Ring 6,
             80805, M\"unchen, Germany} 
\author{\large Cumrun Vafa}
\affiliation{Jefferson Physical Laboratory, Harvard University, Cambridge, MA 02138, USA}

\begin{abstract}
\vspace{0.5cm}
\noindent
We study aspects of anti-de Sitter space in the context of the Swampland. In particular, we conjecture that the near-flat limit of pure AdS belongs to the Swampland, as it is necessarily accompanied by an infinite tower of light states. The mass of the tower is power-law in the cosmological constant, with a power of $\frac{1}{2}$ for the supersymmetric case. We discuss relations between this behaviour and other Swampland conjectures such as the censorship of an unbounded number of massless fields, and the refined de Sitter conjecture. Moreover, we propose that changes to the AdS radius have an interpretation in terms of a generalised distance conjecture which associates a distance to variations of all fields. In this framework, we argue that the distance to the $\Lambda \rightarrow 0$ limit of AdS is infinite, leading to the light tower of states. We also discuss implications of the conjecture for de Sitter space.
\vspace{1cm}
\end{abstract}

\maketitle

\section{Introduction}

The Swampland program aims to distinguish effective field theories that can be completed to quantum gravity in the ultraviolet from those which cannot \cite{Vafa:2005ui} (see \cite{Brennan:2017rbf,Palti:2019pca} for reviews). An interesting aspect of this are considerations regarding the behaviour of effective theories with cosmological constants. String theory is particularly useful in this regard as an explicit theory in which it is possible to calculate the vacuum energy for many vacua, and determine its relation to other features in the theory. The case of a positive vacuum energy, leading to de Sitter space in the effective theory, is particularly difficult to realise in string theory, and this motivated the proposal that it may be obstructed in quantum gravity through a lower bound on the gradient of the potential \cite{Obied:2018sgi}. By contrast, anti-de Sitter (AdS) vacua are simple to realise and there are numerous examples. Nonetheless, the best understood of these examples exhibit certain universal features. In this note we discuss such features and propose that they are general properties of realisations of AdS spaces in quantum gravity. 

AdS vacua in string theory come in discrete families where the AdS radius, or equivalently the negative cosmological constant $\Lambda$, varies. A universal feature of well-understood families of vacua is that the $\Lambda \rightarrow 0$ limit is accompanied by an infinite tower of states becoming light. A simple example is the $AdS_5 \times S^5$ vacuum of type IIB string theory, where sending the AdS cosmological constant to zero also sends the volume of the $S^5$ to infinity, causing the tower of Kaluza-Klein modes to become light. We propose  that the  behaviour of the light tower of states is completely general and formulate the following distance conjecture:
\newline
\newline
{\bf AdS Distance Conjecture\;:}
{\it 
Consider quantum gravity on $d$-dimensional AdS space with cosmological constant $\Lambda$. There exists an infinite tower of states with mass scale $m$ which, as $\Lambda \rightarrow 0$, behaves (in Planck units) as
\be
\label{dsdc}
m \sim |\Lambda|^{\alpha}  \;,
\ee 
where $\alpha$ is a positive order-one number.}
\newline

The AdS Distance Conjecture (ADC) holds for all known backgrounds of string or M-theory which are understood as 10 or 11-dimensional solutions. A common aspect of the solutions is that they actually have $\alpha=\frac12$. We propose that this would be the case for any supersymmetric AdS vacua.
\newline
\newline
{\bf Strong AdS Distance Conjecture\;:}
{\it 
For supersymmetric AdS vacua $\alpha=\frac12$.  
\newline
}

It has been suggested that it is not possible in quantum gravity to induce a parametric gap in AdS between the AdS scale and all massive states \cite{Gautason:2015tig,Gautason:2018gln}. This is a consequence of the Strong ADC, but is generally weaker. The Strong ADC is satisfied in string theory through Kaluza-Klein modes, and therefore amounts to the statement that there is no separation of scales between the AdS radius and internal extra-dimensional radius. This lack of separation of scales was appreciated already in the early supergravity literature,  see for example \cite{Duff:1986hr}, during the flux compactification era, see for example \cite{Douglas:2006es}, and studied continuously until the present, see \cite{Palti:2019pca} for a recent review of these issues. It has a natural interpretation in terms of a statement on the dual CFT: that there is no parametrically controlled gap between the dimensions of a finite number of operators and an infinite number. See, for example, \cite{Gautason:2018gln,Conlon:2018vov} for a discussion of this in recent papers.\footnote{Moreover it can be shown that attempts to shrink the internal spaces such as $S^5$ or $S^7$ by quotienting by a large discrete group, will not eliminate the light KK modes \cite{CJVY}.}  This dual CFT argument should hold for supersymmetric cases at least, as it was proposed in \cite{Ooguri:2016pdq}  that non-supersymmetric AdS always exhibits an instability, which suggests at least the strong ADC.

The aim of this note is to state the conjectures explicitly, and to study their connection to other Swampland conjectures. In particular, we will develop connections to the de Sitter conjecture in section \ref{sec:refds}, to the distance conjecture in section \ref{sec:gdc} (this connection being responsible for the name of the Anti-de Sitter Distance Conjecture and will form the bulk of the note), and to the conjecture that there is a bound on the number of massless modes in quantum gravity in section \ref{sec:boundmass}. We will then discuss the relation to known string vacua in section \ref{sec:intst}, and finally discuss the de Sitter case in section \ref{sec:ds}.

\section{Refined de Sitter conjecture}
\label{sec:refds}

A direct connection between the Strong ADC and the refined de Sitter conjecture \cite{Ooguri:2018wrx,Garg:2018reu} can be noted. The refined de Sitter conjecture proposes a bound on the derivatives of the potential  
\begin{equation}
\label{rdsc1}
\left|\nabla V \right| \geq c V\;,
\end{equation}
or
\begin{equation}
\label{rdsc2}
\mathrm{min}\left(\nabla_i \nabla_j V \right) \leq -c' V\;.
\end{equation}
With $c$ and $c'$ positive order one constants.
We propose that condition (\ref{rdsc1}) may be naturally promoted to the stronger statement 
\be
\left|\nabla V \right|^2 \geq c^2 V^2\;.
\ee
In that case a minimum in AdS space would require satisfying the second condition (\ref{rdsc2}). As was already noted in \cite{Gautason:2018gln}, that implies that the mass of the lightest state is bounded from above by $\left|c'\Lambda\right|^{\frac12}$. Such bounds on the mass of a state are familiar from statements such as the Weak Gravity Conjecture \cite{ArkaniHamed:2006dz}. It has been proposed a number of times that in fact the WGC bounds should be satisfied by an infinite tower of states \cite{Heidenreich:2015nta,Klaewer:2016kiy,Montero:2016tif,Heidenreich:2016aqi,Andriolo:2018lvp,Grimm:2018ohb}. The ADC is then naturally a tower version of the (slightly adjusted) de Sitter conjecture applied to AdS space. This perspective on the ADC also suggests that for general AdS vacua, not only supersymmetric, $\alpha \geq \frac12$. Implying the tower mass scale $m \sim \left|\Lambda\right|^{\alpha} \leq \left| \Lambda \right|^{\frac12}$ as $\Lambda \rightarrow 0$.

\section{Distance Conjecture}
\label{sec:gdc}

The ADC refers to an infinite tower of states which become massless. Such a tower is familiar from other Swampland criteria, see the review \cite{Palti:2019pca}. In particular, the distance conjecture \cite{Ooguri:2006in} states that such an infinite tower of states appears at infinite distances in scalar moduli space. We consider a theory with gravity and massless (or light) real scalar fields $\phi^i$, which have a field-space metric $p_{ij}(\phi)$ such that their kinetic term in the Einstein frame is
\be
{\cal L}_{\mathrm{kin}} = -p_{ij}(\phi)\partial \phi^i \partial \phi^j  \;.
\ee
Then we consider a geodesic path $\gamma$ in field space, which we can choose to parametrise by $\tau$ ranging from an initial $\tau_i$ to a final $\tau_f$,   so that the proper distance $\Delta\phi$ along the path is
\be
\Delta \phi = \int_{\tau_i}^{\tau_f} \left(p_{ij}\frac{\partial \phi^i}{\partial \tau} \frac{\partial \phi^j}{\partial \tau} \right)^{\frac12} d\tau \;.
\label{dissc}
\ee
Then the conjecture states that there must be an infinite tower of states with mass scale $m\left(\tau\right)$ such that 
\be
m\left(\tau_f\right) \sim m\left(\tau_i\right) e^{-\alpha {\Delta \phi}}\;.
\label{dsc}
\ee 
Here, $\alpha$ is conjectured to always be an ${\cal O}(1)$ positive parameter (see \cite{Baume:2016psm,Klaewer:2016kiy} for discussion of this aspect), and (\ref{dsc}) is expected to hold for $\Delta \phi > {\cal O}(1)$ in Planck units becoming sharper as $\Delta \phi \rightarrow \infty$. The distance conjecture has been proposed to hold even for fields with a potential \cite{Baume:2016psm,Klaewer:2016kiy}, and has been significantly tested in string theory, see for example \cite{Cecotti:2015wqa,Palti:2015xra,Baume:2016psm,Klaewer:2016kiy,Valenzuela:2016yny,Blumenhagen:2017cxt,Valenzuela:2017bvg,Palti:2017elp,Hebecker:2017lxm,Cicoli:2018tcq,Grimm:2018ohb,Blumenhagen:2018nts,Blumenhagen:2018hsh,Lee:2018urn,Lee:2018spm,Grimm:2018cpv,Kim:2018vgz,Buratti:2018xjt,Hebecker:2018fln,Corvilain:2018lgw,Joshi:2019nzi,Scalisi:2018eaz,Lee:2019tst,Blumenhagen:2019qcg,Marchesano:2019ifh,Font:2019cxq,Lee:2019xtm,Grimm:2019wtx,Erkinger:2019umg}.\footnote{There is also a proposal that it is a consequence of the emergent nature of scalar fields in quantum gravity \cite{Grimm:2018ohb,Heidenreich:2018kpg,Ooguri:2018wrx,Palti:2019pca}.}

The distance conjecture is capturing bounds on how much scalar fields can change in effective theories coming from string theory. However, given that often (but not always) the ultraviolet origin of scalar fields is in variations of metric and anti-symmetric gauge fields in the internal space, it is natural to ask how the distance conjecture can be more generally formulated to apply to all fields.  In particular, we look for a formulation so that that the scalar field distance conjecture (\ref{dsc}) would be a consequence of a Generalized Distance Conjecture applied to all fields.  

Any dynamical field has a natural metric on its variations induced by its kinetic terms. So given a field with $n$ spacetime indices, ${\cal O}_{M_1...M_n}$, we can write its kinetic term as
\be
{\cal L}_{\mathrm{kin}} = -G^{M_1...M_nN_1...N_n} D {\cal O}_{M_1...M_n}D{\cal O}_{N_1...N_n}   \;,
\ee
where $D$ is some differential operator. $G$ is then a metric on the field space of ${\cal O}$. We can then associate a distance in the field space $\Delta {\cal O}$ as
\be
\Delta {\cal O} = \int_{\tau_i}^{\tau_f} \left( G \frac{\partial {\cal O}}{\partial \tau} \frac{\partial {\cal O}}{\partial \tau}   \right)^{\frac12}d\tau \;.
\label{insope}
\ee
To make sense out of this distance, we have to explain what we do when the vacuum expectation values of fields ${\cal O}$ depend on the spacetime coordinates, as in general would be the case for the internal dimensions. Naively one would be led to simply integrate over the space.  However, taking the original case of scalar fields and integrating over spacetime leads to infinity.  Instead we can view this as the average value of the variation over the spacetime, in which case the volume of space time cancels out.  In other words we replace (\ref{insope}) by
\be
\Delta {\cal O} = \int_{\tau_i}^{\tau_f} \left(\langle G \frac{\partial {\cal O}}{\partial \tau} \frac{\partial {\cal O}}{\partial \tau}   \rangle \right)^{\frac12}d\tau \;,
\label{insop}
\ee
where 
$$\langle G \frac{\partial {\cal O}}{\partial \tau} \frac{\partial {\cal O}}{\partial \tau}   \rangle ={1\over V_M} \int _M G \frac{\partial {\cal O}}{\partial \tau} \frac{\partial {\cal O}}{\partial \tau} \;, $$ 
with $M$ is the spacetime manifold, with volume $V_M$.  If we imagine $M=S\times Y$ where $Y$ is the compact internal space (which could be empty) and $S$ is the homogeneous Einstein space (Minkowski, AdS or dS), this leads using the homogeneity to 
$$\langle G \frac{\partial {\cal O}}{\partial \tau} \frac{\partial {\cal O}}{\partial \tau}   \rangle ={1\over V_Y} \int _Y G \frac{\partial {\cal O}}{\partial \tau} \frac{\partial {\cal O}}{\partial \tau}$$
If the scalar field kinetic terms come from variations of the internal fields, such as metric and gauge fields, this is exactly how we would compute the metric on the scalar field space, upon compactifying the theory on $Y$. Note that dividing by $1/V_Y$ has the effect of going to the Einstein frame, because the kinetic term for the scalars and the Einstein term scale the same way under Weyl rescalings and the Einstein term has an extra factor of $V_Y$ which needs to be gotten rid of by Weyl rescaling.

In fact, this is not quite general. Going to the Einstein frame only amounts to dividing by the volume in the case where the variations do not modify the volume $V_M$. When they do modify it there is also a correction to the kinetic terms coming from the transformation. More precisely, if we consider a compactification of type $M = S_d \times Y_k$, where the subscripts denote the real dimensions of the external and internal spaces, then the Einstein frame on $S_d$ leads to the kinetic terms for the volume variation of $Y_k$, which take the form
\be
-k^2\left[  \frac{\left(d-1\right)}{\left(d-2\right)} - \frac{\left(k-1\right)}{k} \right] \left( \partial \tau \right)^2 \;,
\label{Weylrescvol}
\ee 
where we parameterised the volume of $Y_k$ as $\left(V_Y\right)^{\frac{1}{k}} = e^{\tau}$. We therefore can decompose the possible field variations to those which preserve the volume, and the volume itself, the latter having a metric on its variations (\ref{Weylrescvol}). This is true for volume variations of $Y_k$, we can also consider volume variations of $S_d$, and we will discuss this in section \ref{sec:weylvar}. 

Note that if there are many different types of fields ${\cal O}_{(i)}$ which vary with some parameter of the possible solutions to the equations of motion, then the distance is the total distances for all the fields. Let us state this prescription explicitly. We consider a family of (static) solutions to the equations of motion, not necessarily degenerate in energy, specified by a parameter $\tau$. We then promote this parameter to have some weak external spacetime dependence, time only is sufficient, and extract the kinetic terms of that parameter $-K\left(\tau\right) \left(\partial \tau \right)^2$. These will receive contribution from all the fields which vary with $\tau$, so
\be
K\left(\tau\right) = \sum_i \langle G_{(i)} \frac{\partial {\cal O}_{(i)}}{\partial \tau} \frac{\partial {\cal O}_{(i)}}{\partial \tau}   \rangle \;.
\label{Kdef}
\ee  
The total distance is then
\be
\Delta = \int_{\tau_i}^{\tau_f}\sqrt{\left|K\right|} d\tau \;. 
\ee

We are now ready to state the generalized distance conjecture:  Consider the non-compact space to be an Einstein space, i.e. AdS, Minkowski, or dS (if it exists). Then for large distance variation in fields, we get a light tower of states in the Einstein frame of the external effective field theory, whose mass scale in Planck units is given by
\be
m \sim e^{-\alpha \Delta }\;,
\label{gendisc}
\ee
where $\alpha \sim {\cal O}(1)$.

\subsection{Example: transverse traceless metric variations}

An interesting application of the generalised distance conjecture is to the metric, so utilising a notion of distance in the space of metrics. The procedure outlined in section \ref{sec:gdc} suggests to read off the metric on the space through the kinetic terms. We consider expanding the metric as a perturbation about an arbitrary background $g^0_{MN}$
\be
g_{MN} = g^0_{MN} + \delta g_{MN} \;.
\label{metvar}
\ee
We first restrict to the case of transverse and traceless variations
\be
\nabla^{M} \delta g_{MN} = 0 \;,\;\; \delta g^M_{\;\;M} = 0 \;.
\label{trtr}
\ee
Then the kinetic terms, coming from the expansion of the Ricci scalar, yield the expression for the metric on the field space of such metric deformations. If we consider an arbitrary path, labelled with parameter $\tau$, through transverse-traceless metric space, then the associated distance is  
\be
\Delta = c \int_{\tau_i}^{\tau_f}  \left( \frac{1}{V_M} \int_M \sqrt{g} g^{MN} g^{OP} \frac{\partial g_{MO}}{\partial \tau} \frac{\partial g_{NP}}{\partial \tau}   \right)^{\frac12}d\tau \;,
\label{dismetr}
\ee
where $c \sim {\cal O}(1)$ is a constant depending on the dimension of $M$. This is a well-known notion of distance in metric spaces, see for example \cite{PhysRev.160.1113}, and has also been studied from a pure mathematics perspective, including the notion of geodesics on the space, see for example \cite{1992math......1259G}.

If we apply the general formula (\ref{dismetr}) to the case of a Calabi-Yau compactification $M = \mathrm{Mink}_4 \times \mathrm{CY}_6$, for metric variations that are zero modes, then it reduces to nothing but the scalar distance (\ref{dissc}) on the moduli spaces \cite{Candelas:1990pi}. 

%Note that if we dropped the constraints (\ref{trtr}) then the expression for the distance (\ref{dismetr}) would be modified to ensure that the variations are orthogonal to diffeomorphisms \cite{Douglas:2008jx,EP}. 

\subsection{External Weyl rescalings}
\label{sec:weylvar}

We are particularly interested in distances in metric space associated to variations of the external metric that are Weyl rescalings. Extracting this distance is subtle because it is not clear how to implement the analogue of the Einstein frame condition when the external space volume is itself changing. 

On the one hand, up to order one factors, the metric on such variations is simple to determine since we already extracted it in the external Weyl transformation contribution of the first term in (\ref{Weylrescvol}).\footnote{Another example which is purely four-dimensional is $R+R^2$, or also pure $R^2$, gravity which propagates a physical scalar field whose kinetic terms come purely from the external Weyl transformation.}  So we strongly expect that under an external Weyl variation of the external space
\be
\tilde{g}_{\mu\nu} = e^{2\tau} g_{\mu\nu}\;,
\label{weytra}
\ee
the canonically normalised field for measuring the distance is, up to order one constants, $\tau$.

On the other hand, it is not clear to us how to exactly determine the factor, and sign, of the distance contribution. One natural suggestion is to perform a type of dimensional regularisation by adding a fictitious $d'$-dimensional external space, go to the Einstein frame of that space, and then send $d' \rightarrow 0$. This would yield a kinetic term for the external $\tau$ which is given by replacing $k \rightarrow d$ and $d \rightarrow 0$ in (\ref{Weylrescvol}). This leads to a negative contribution to the distance going as
\be
K_{\left(\tau\right)} = -d^2 \left[ \frac{\left(d-1\right)}{d} - \frac12\right] \;.
\label{Kdistau}
\ee
Another possibility is to perform the external Weyl rescaling, extract the induced kinetic term, and then drop the overall prefactor. This would lead again to a negative distance contribution of $K_{\left(\tau\right)} = -\left(d-2\right)\left(d-1\right)$. 

It is not completely clear what to make of the negative distance contribution. 
%At a fundamental level it relates to the question of whether the notion of distance can be applied to directions in field space which are not local propagating degrees of freedom. Specifically, in four dimensions, the metric only leads to two local graviton polarization modes which are transverse and traceless. The Weyl deformation is not traceless, but is clearly is a physical variation of the metric field. We therefore believe that it should have a distance associated to it. 
%
%Indeed, if we consider a compactification scenario, say $M = \mathrm{Mink}_4 \times \mathrm{CY}_6$, then the volume of the Calabi-Yau is also not traceless, but leads to a physical scalar field satisfying the distance conjecture in the effective four-dimensional theory. More precisely, in order to extract the kinetic terms for such a variation it must be combined with a Weyl variation of the external metric (transforming to the Einstein frame).  While the direct dimensional reduction of the Ricci scalar leads to negative kinetic terms, the final kinetic terms (and distance) after going to the Einstein frame are positive.
%
We propose that the negative distance contribution of the external Weyl factor is physical, and its negative sign is also meaningful. However, it will only be one negative contribution to a total variation of all the fields along a path which solves the equations of motion, as in (\ref{Kdef}). We propose that the total distance will then always be positive. 

\subsection{Application to AdS}
\label{sec:apads}

We now apply the generalised distance conjecture, developed in the previous section, to AdS vacua. In global coordinates the $AdS_d$ metric takes the form
\be
ds^2 = e^{2\tau} \left( - \left(\cosh \rho\right)^2dt^2 + d\rho^2 + \left(\sinh \rho\right)^2 d \Omega_{d-2}^2 \right) \;.
\ee
With the radius being related to the cosmological constant as
\be
\Lambda = -\frac12 \left(d-1\right)\left(d-2\right) e^{-2\tau} \;.
\ee
We consider a family of AdS vacua parametrised by a variation of the cosmological constant from $\Lambda_i$ to $\Lambda_f$. In string theory, such a family of vacua would be separated by a distance in field space as defined by the generalised distance conjecture. Irrespective of the details of the variations of all the fields, we can extract a universal contribution to the distance from the variation of the external AdS metric. As discussed in section \ref{sec:weylvar}, we are not certain of the exact factor for this distance, but in the natural case it yields a contribution $K_{\left(\tau\right)}$ which is negative and order one. So the total distance takes the form
\be
\Delta = -\frac12 \int_{\Lambda_i}^{\Lambda_f}  \left| K_{\left(\tau\right)}  + K' \right|^{\frac12} d \log \Lambda \;.
\ee
Here $K'$ denotes the contribution to the distance from the variation of all the fields other than the external metric. The contribution from any fields which have positive kinetic terms will always be positive, and so we expect in general $K' > 0$. 

We now propose that the contribution from the other fields $K'$, will be at least as large as that of the external metric, to make the factor inside the bracket positive.\footnote{Note that the variation of other fields will therefore also be at infinite distance. In the case of such variations of fluxes, it is natural to expect that $\left(d-2\right)$ branes will be relevant objects.} Further, we propose that the resulting total contribution will still be of order the external metric contribution, so 
\be
\left|K_{\left(\tau\right)}  + K' \right|^{\frac12} \sim {\cal O}(1) \;.
\label{notuning}
\ee   
This proposal can then be combined with the generalised distance conjecture (\ref{gendisc}) to predict that there must be an infinite tower of states whose mass scale behaves as
\be
m\left(\Lambda_f\right) \sim m\left(\Lambda_i\right) \left(\frac{\Lambda_f}{\Lambda_i} \right)^{\alpha}\;,
\ee
with $\alpha \sim {\cal O}(1)$. We can consider the initial value of $\Lambda_i$ to be near the Planck scale, and similarly for the mass scale $m\left(\Lambda_i\right)$. This then gives
\be
m\left(\Lambda\right) \sim M_p \left(\frac{\Lambda}{M_p^2} \right)^{\alpha}\;,
\label{distinmp}
\ee
which, in Planck units, is the AdS Distance Conjecture (\ref{dsdc}). 

It is important to emphasise that our proposal, or assumption, that there is no fine tuned cancellation between the infinite negative and positive contributions in (\ref{notuning}), translates in the ADC to the $\alpha \sim {\cal O}\left(1\right)$ proposal. It would be good to test how small $\alpha$ could be made in string theory examples to make this aspect sharper.

\section{Bounded Massless Matter}
\label{sec:boundmass}

When the tower of light states of the strong AdS distance conjecture are Kaluza-Klein states, it amounts to the statement of the absence of separation of scales between the external and internal spaces. This may be related to another Swampland conjecture which is that there should not be an unbounded number of spacetime massless states in quantum gravity \cite{Vafa:2005ui} (see also \cite{Acharya:2006zw,Heckman:2019bzm}). This, at first, appears to be violated in AdS examples.  More precisely, string theory on backgrounds which include an $\mathrm{AdS}_d$ factor permit an unbounded number of massless fields which propagate completely in the $\mathrm{AdS}_d$.   An example of this is M-theory on $\mathrm{AdS}_7 \times S^4/Z_k$. This has been extensively studied, see for example 
\cite{Hanany:1997sa,Brunner:1997gf,Ahn:1998pb,Hanany:1997gh,Ferrara:1998vf,Apruzzi:2013yva,Gaiotto:2014lca,DelZotto:2014hpa,Louis:2015mka} 
and is dual to ${\cal N}=(1,0)$ superconformal theory of M5 branes probing $A_{k-1}$ singularity.   This theory has $SU(k)\times SU(k)$ global symmetry and so its gravity dual should have $SU(k)\times SU(k)$ gauge symmetry.  Indeed the $\mathbb{Z}_k$ orbifolding induces two $A_{k-1}$ singularities at each pole of $S^4$. We then obtain two $SU(k)$ gauge symmetries, and are free to take $k \rightarrow \infty$ to obtain an unbounded number of massless modes propagating in $AdS_7$.

However, this is actually compatible with a bound on spacetime massless matter because in the absence of separation of scales we should view the $S^4$ as part of the space because it cannot be made parametrically small compared to $AdS_7$ radius.  The massless matter is not actually propagating in the full spacetime. There is no sense in which this is an $\mathrm{AdS}_7$ spacetime, and so the massless matter is localised on a codimension 4 defect in a higher dimensional spacetime. There is no expectation that an infinite number of massless modes living on a defect is not allowed in quantum gravity. The infinite number of massless fields may therefore be an example of quantum gravity censoring a limit $\mathrm{AdS}_d \rightarrow \mathrm{Mink}_d$. Rather, it suggests that one must always have a limit such as $\mathrm{AdS}_d \times Y_p \rightarrow \mathrm{Mink}_{d+p}$, where $Y_p$ is a $p$-dimensional internal space. Or in other words, there is no nearly-flat pure AdS space in quantum gravity, only spaces which contain an AdS factor.

\section{The ADC in string theory}
\label{sec:intst}

As mentioned previously, to our knowledge, the ADC and its strong version are satisfied in any known vacuum of string theory for which we have a well-understood higher-dimensional solution. 
%There are some general results regarding the strong version which suggest that it should be satisfied at least in any vacuum of string theory which admits a consistent truncation to a finite number of fields. 
%Indeed, this was proposed as a conjecture in \cite{Gauntlett:2007ma}. 
There are some general results regarding this.\footnote{Note that in \cite{Banks:2016xsj} a holographic argument for no separation of scales was presented.} In \cite{Gautason:2015tig} it was argued that separation of scales is not possible in pure supergravity settings, and in string theory settings should be possible only if there are non-vanishing spatial gradients for scalar fields in the internal manifold. This matches well known results in type IIA compactifications where the Calabi-Yau compactification admits a separation of scales \cite{DeWolfe:2005uu}, but is not an exact solution due to localised sources. Backreacting these sources must induce, at least, spatial gradients such as warp factors and dilaton gradients. The key open question in this setting is whether the fully backreacted solution still allows for separation of scales. This may be connected to the existence of a consistent truncation to a finite number of fields, since such truncations are typically associated to some coset structure on the manifold, and it is well known that gradients for scalar fields are incompatible with a coset structure (see, for example, \cite{Palti:2006yz} for a discussion of this). Indeed, in \cite{Gauntlett:2007ma} it was mentioned that in general there is no separation of scales in consistent truncations, and was also conjectured that all supersymmetric AdS supergravity vacua admit a consistent truncation.   
A connection between separation of scales, and the absence of a consistent truncation, would imply that finding fully controlled solution of string theory exhibiting parametric separation of scales, if they exist, will be a difficult task at best.\footnote{This is also related to the result in \cite{Micu:2007rd} that geometric compactifications of string theory to supersymmetric Minkowski vacua with no massless fields are not possible. It was argued in \cite{Palti:2007pm} that it may be possible for non-geometric compactifications, but it is unclear if these are under sufficient control to be trustable.} 

If we consider proposals for string vacua which do not yet have a well understood uplift to a higher dimensional supergravity solution, so which utilise intrinsically four-dimensional arguments, then one may violate the conjectures. More precisely, consider three of the leading scenarios, that is the Kachru-Kallosh-Linde-Trivedi (KKLT) scenario \cite{Kachru:2003aw}, the Large Volume Scenario (LVS) in type IIB string theory \cite{Balasubramanian:2005zx}, and the type IIA Calabi-Yau scenarios (DGKT) \cite{DeWolfe:2005uu}.\footnote{There are also type IIA realisations of the LVS scenario \cite{Palti:2008mg}, which behave with respect to the conjectures in this paper the same as the type IIB LVS.} \footnote{It would be interesting to extend tests to very general classes of lower-dimensional supergravity vacua, for an analysis of such AdS vacua see for example \cite{Louis:2016tnz,Lust:2017aqj}.} Then the LVS scenario satisfies both the weak and strong versions of the AdS distance conjectures, the latter due to not being supersymmetric. The type IIA DGKT scenarios satisfy the weaker conjecture (\ref{dsdc}), but violate the strong version. The KKLT scenario does not explicitly satisfy either of the conjectures, but may do so implicitly. By this we mean that assuming the KKLT constant $W_0$ can be tuned arbitrarily small by fluxes, the Kaluza-Klein towers associated to four-cycles would only scale logarithmically with the resulting cosmological constant $m^{-1}_{t_{KK}} \sim -\log \left|\Lambda\right|$. However, it may still be that (\ref{dsdc}) is satisfied by a different tower of states, in particular there are towers whose mass scale is controlled by the dilaton and by the complex-structure moduli.\footnote{Such towers may also be realised by KK or string modes in induced long throats near conifold loci, for example as discussed in  \cite{Bena:2018fqc,Blumenhagen:2019qcg}.}
Whether such towers satisfy (\ref{dsdc}) or not depends on doing explicit calculations of the values of the complex-structure moduli in vacua leading to very small $W_0$. So if KKLT is compatible with (\ref{dsdc}), then it is implicitly so, either through light towers of states, or through a bound on how small $W_0$ can actually be made in string theory.

Another relevant point is that in the case of moduli spaces the different vacua points are connected by a continuous set of vacua. This is not the case for AdS vacua in string theory. Nonetheless, we expect that the notion of the distance should not depend on the existence of such interpolating vacua. The Weak Gravity Conjecture provides a useful analogy: the tower of states at $g M_p$ does not depend on the existence of vacua continuously connecting the different values of the gauge coupling $g$.

\section{Implications for de Sitter}
\label{sec:ds}

In sections \ref{sec:gdc} and \ref{sec:apads} we introduced the generalised distance conjecture and showed that in this context the $\Lambda \rightarrow 0$ limit of AdS space is at infinite distance. In fact, the same analysis applies equally to de Sitter space for a positive $\Lambda$, so we expect a tower of states at mass scale $m \sim \Lambda^{\alpha}$.  In this picture we therefore have three branches, AdS, Minkowski, and dS, which are separated by infinite distance in metric space.\footnote{This is in analogy to having a particle with negative, zero, or positive charge, such that these branches are separated by infinite distance in the field space of fields which control the gauge coupling.} This would suggest that there is no effective field theory with a finite number of fields which can have families of vacua interpolating from one branch to the other. So passing or approaching $\Lambda =0$ will lead to an unbounded number of massless modes. In particular, this means that if we have a theory holographically dual to AdS, then it is not possible to describe a dS vacuum as a state in that theory.\footnote{For studies in this direction see for example \cite{Freivogel:2005qh}.}  This observation shows that the generalized distance conjecture lends further support to the dS swampland conjecture, suggesting that meta-stable dS vacua indeed do not exist.

There is also an interesting insight into the value of $\alpha$ in the de Sitter case. The towers of states which appear in string theory typically involve particles with spin greater than or equal to 2. For example, Kaluza-Klein towers will always have massive graviton modes. It is known that in de Sitter space, or in sufficiently quasi-de Sitter space, there is a bound on the mass of a particle with spin 2 or higher relative to the cosmological constant. The Higuchi bound states \cite{Higuchi:1986py} 
\be
m_{\mathrm{Spin \geq 2}} \gtrsim \Lambda^{\frac12} \;.
\ee
This implies that for $\Lambda \ll 1$ we must have $\alpha \leq \frac12$. It is interesting that the dS case suggests the opposite direction bound to the AdS case discussed in section \ref{sec:refds}. 

Note that such a change in the bound is unusual for Swampland conjectures, which usually appear at the quadratic level and are not sensitive to sign changes. Given this, it is also interesting to consider the possibility that $\alpha \geq \frac12$ could be universal to both dS and AdS spaces. If that is so, then it would imply that the late-time acceleration of our universe must depart sufficiently from de Sitter so as to modify the Higuchi bound sufficiently to be compatible with a lower bound on $\alpha$.

For our universe, if the dS conjecture in \cite{Obied:2018sgi,Ooguri:2018wrx} were false, and moreover if we live in a dS space with a small cosmological constant, then our proposal predicts a light tower of states 
\be
m \sim 10^{-120 \alpha} M_p \;.
\label{dsm}
\ee
Since the states need only couple gravitationally to the Standard Model, it is natural to expect that the observational bounds on the tower are similar to those of long-range modifications of gravity $m \gtrsim 10^{-30} eV$, implying $\alpha < \frac12$. However, we have not performed a detailed analysis, and the fact that it is a tower of states rather than a single scalar field can have interesting cosmological consequences. It is interesting that this is relating the smallness of the cosmological constant with a hierarchic lightness of a tower of states which could be part of the visible or dark sector.

A similar conclusion, that a tower of states may become light in our cosmology, has already been suggested in the context of the dS swampland conjecture as a consequence of quintessence field rolling a large distance \cite{Agrawal:2018own}. Assuming this tower is in the dark sector it leads to an interesting phenomenology \cite{AOV}.  Our results therefore suggest that regardless of whether we assume the validity of the dS swampland conjecture, we are led to expect a tower of light states in the current universe! It is interesting to note that this possibly relates the hierarchy problem with the smallness of dark energy in our universe.
\newline
\newline
{\bf Comment:} While this note was in final preparation, \cite{Alday:2019qrf} appeared which has some overlap with this topic.

\vspace{10px}
{\bf Acknowledgements}
\noindent

The work of D.L. is supported by the Origins Excellence Cluster.  The research of C.V. is supported in part by the NSF grant
PHY-1719924 and by a grant from the Simons Foundation (602883, CV). We would like to thank P. Agrawal, D. Jafferis, J. Maldacena and G. Obied for discussions. D.L. and E.P. would like to thank Harvard University, where this project was initiated, for hospitality.

\appendix

\bibliography{adsswamp}

\begin{thebibliography}{80}%
\makeatletter
\providecommand \@ifxundefined [1]{%
 \@ifx{#1\undefined}
}%
\providecommand \@ifnum [1]{%
 \ifnum #1\expandafter \@firstoftwo
 \else \expandafter \@secondoftwo
 \fi
}%
\providecommand \@ifx [1]{%
 \ifx #1\expandafter \@firstoftwo
 \else \expandafter \@secondoftwo
 \fi
}%
\providecommand \natexlab [1]{#1}%
\providecommand \enquote  [1]{``#1''}%
\providecommand \bibnamefont  [1]{#1}%
\providecommand \bibfnamefont [1]{#1}%
\providecommand \citenamefont [1]{#1}%
\providecommand \href@noop [0]{\@secondoftwo}%
\providecommand \href [0]{\begingroup \@sanitize@url \@href}%
\providecommand \@href[1]{\@@startlink{#1}\@@href}%
\providecommand \@@href[1]{\endgroup#1\@@endlink}%
\providecommand \@sanitize@url [0]{\catcode `\\12\catcode `\$12\catcode
  `\&12\catcode `\#12\catcode `\^12\catcode `\_12\catcode `\%12\relax}%
\providecommand \@@startlink[1]{}%
\providecommand \@@endlink[0]{}%
\providecommand \url  [0]{\begingroup\@sanitize@url \@url }%
\providecommand \@url [1]{\endgroup\@href {#1}{\urlprefix }}%
\providecommand \urlprefix  [0]{URL }%
\providecommand \Eprint [0]{\href }%
\providecommand \doibase [0]{http://dx.doi.org/}%
\providecommand \selectlanguage [0]{\@gobble}%
\providecommand \bibinfo  [0]{\@secondoftwo}%
\providecommand \bibfield  [0]{\@secondoftwo}%
\providecommand \translation [1]{[#1]}%
\providecommand \BibitemOpen [0]{}%
\providecommand \bibitemStop [0]{}%
\providecommand \bibitemNoStop [0]{.\EOS\space}%
\providecommand \EOS [0]{\spacefactor3000\relax}%
\providecommand \BibitemShut  [1]{\csname bibitem#1\endcsname}%
\let\auto@bib@innerbib\@empty
%</preamble>
\bibitem [{\citenamefont {Vafa}(2005)}]{Vafa:2005ui}%
  \BibitemOpen
  \bibfield  {author} {\bibinfo {author} {\bibfnamefont {C.}~\bibnamefont
  {Vafa}},\ }\href@noop {} {\  (\bibinfo {year} {2005})},\ \Eprint
  {http://arxiv.org/abs/hep-th/0509212} {arXiv:hep-th/0509212 [hep-th]}
  \BibitemShut {NoStop}%
%%CITATION = HEP-TH/0509212;%%
\bibitem [{\citenamefont {Brennan}\ \emph {et~al.}(2017)\citenamefont
  {Brennan}, \citenamefont {Carta},\ and\ \citenamefont
  {Vafa}}]{Brennan:2017rbf}%
  \BibitemOpen
  \bibfield  {author} {\bibinfo {author} {\bibfnamefont {T.~D.}\ \bibnamefont
  {Brennan}}, \bibinfo {author} {\bibfnamefont {F.}~\bibnamefont {Carta}}, \
  and\ \bibinfo {author} {\bibfnamefont {C.}~\bibnamefont {Vafa}},\ }\bibfield
  {booktitle} {\emph {\bibinfo {booktitle} {{Proceedings, Theoretical Advanced
  Study Institute in Elementary Particle Physics: Physics at the Fundamental
  Frontier (TASI 2017): Boulder, CO, USA, June 5-30, 2017}}},\ }\href {\doibase
  10.22323/1.305.0015} {\bibfield  {journal} {\bibinfo  {journal} {PoS}\
  }\textbf {\bibinfo {volume} {TASI2017}},\ \bibinfo {pages} {015} (\bibinfo
  {year} {2017})},\ \Eprint {http://arxiv.org/abs/1711.00864} {arXiv:1711.00864
  [hep-th]} \BibitemShut {NoStop}%
%%CITATION = ARXIV:1711.00864;%%
\bibitem [{\citenamefont {Palti}(2019)}]{Palti:2019pca}%
  \BibitemOpen
  \bibfield  {author} {\bibinfo {author} {\bibfnamefont {E.}~\bibnamefont
  {Palti}}\ }(\bibinfo {year} {2019})\ \Eprint
  {http://arxiv.org/abs/1903.06239} {arXiv:1903.06239 [hep-th]} \BibitemShut
  {NoStop}%
%%CITATION = ARXIV:1903.06239;%%
\bibitem [{\citenamefont {Obied}\ \emph {et~al.}(2018)\citenamefont {Obied},
  \citenamefont {Ooguri}, \citenamefont {Spodyneiko},\ and\ \citenamefont
  {Vafa}}]{Obied:2018sgi}%
  \BibitemOpen
  \bibfield  {author} {\bibinfo {author} {\bibfnamefont {G.}~\bibnamefont
  {Obied}}, \bibinfo {author} {\bibfnamefont {H.}~\bibnamefont {Ooguri}},
  \bibinfo {author} {\bibfnamefont {L.}~\bibnamefont {Spodyneiko}}, \ and\
  \bibinfo {author} {\bibfnamefont {C.}~\bibnamefont {Vafa}},\ }\href@noop {}
  {\  (\bibinfo {year} {2018})},\ \Eprint {http://arxiv.org/abs/1806.08362}
  {arXiv:1806.08362 [hep-th]} \BibitemShut {NoStop}%
%%CITATION = ARXIV:1806.08362;%%
\bibitem [{\citenamefont {Gautason}\ \emph {et~al.}(2016)\citenamefont
  {Gautason}, \citenamefont {Schillo}, \citenamefont {Van~Riet},\ and\
  \citenamefont {Williams}}]{Gautason:2015tig}%
  \BibitemOpen
  \bibfield  {author} {\bibinfo {author} {\bibfnamefont {F.~F.}\ \bibnamefont
  {Gautason}}, \bibinfo {author} {\bibfnamefont {M.}~\bibnamefont {Schillo}},
  \bibinfo {author} {\bibfnamefont {T.}~\bibnamefont {Van~Riet}}, \ and\
  \bibinfo {author} {\bibfnamefont {M.}~\bibnamefont {Williams}},\ }\href
  {\doibase 10.1007/JHEP03(2016)061} {\bibfield  {journal} {\bibinfo  {journal}
  {JHEP}\ }\textbf {\bibinfo {volume} {03}},\ \bibinfo {pages} {061} (\bibinfo
  {year} {2016})},\ \Eprint {http://arxiv.org/abs/1512.00457} {arXiv:1512.00457
  [hep-th]} \BibitemShut {NoStop}%
%%CITATION = ARXIV:1512.00457;%%
\bibitem [{\citenamefont {Gautason}\ \emph {et~al.}(2019)\citenamefont
  {Gautason}, \citenamefont {Van~Hemelryck},\ and\ \citenamefont
  {Van~Riet}}]{Gautason:2018gln}%
  \BibitemOpen
  \bibfield  {author} {\bibinfo {author} {\bibfnamefont {F.~F.}\ \bibnamefont
  {Gautason}}, \bibinfo {author} {\bibfnamefont {V.}~\bibnamefont
  {Van~Hemelryck}}, \ and\ \bibinfo {author} {\bibfnamefont {T.}~\bibnamefont
  {Van~Riet}},\ }\href {\doibase 10.1002/prop.201800091} {\bibfield  {journal}
  {\bibinfo  {journal} {Fortsch. Phys.}\ }\textbf {\bibinfo {volume} {67}},\
  \bibinfo {pages} {1800091} (\bibinfo {year} {2019})},\ \Eprint
  {http://arxiv.org/abs/1810.08518} {arXiv:1810.08518 [hep-th]} \BibitemShut
  {NoStop}%
%%CITATION = ARXIV:1810.08518;%%
\bibitem [{\citenamefont {Duff}\ \emph {et~al.}(1986)\citenamefont {Duff},
  \citenamefont {Nilsson},\ and\ \citenamefont {Pope}}]{Duff:1986hr}%
  \BibitemOpen
  \bibfield  {author} {\bibinfo {author} {\bibfnamefont {M.~J.}\ \bibnamefont
  {Duff}}, \bibinfo {author} {\bibfnamefont {B.~E.~W.}\ \bibnamefont
  {Nilsson}}, \ and\ \bibinfo {author} {\bibfnamefont {C.~N.}\ \bibnamefont
  {Pope}},\ }\href {\doibase 10.1016/0370-1573(86)90163-8} {\bibfield
  {journal} {\bibinfo  {journal} {Phys. Rept.}\ }\textbf {\bibinfo {volume}
  {130}},\ \bibinfo {pages} {1} (\bibinfo {year} {1986})}\BibitemShut {NoStop}%
%%CITATION = PRPLC,130,1;%%
\bibitem [{\citenamefont {Douglas}\ and\ \citenamefont
  {Kachru}(2007)}]{Douglas:2006es}%
  \BibitemOpen
  \bibfield  {author} {\bibinfo {author} {\bibfnamefont {M.~R.}\ \bibnamefont
  {Douglas}}\ and\ \bibinfo {author} {\bibfnamefont {S.}~\bibnamefont
  {Kachru}},\ }\href {\doibase 10.1103/RevModPhys.79.733} {\bibfield  {journal}
  {\bibinfo  {journal} {Rev. Mod. Phys.}\ }\textbf {\bibinfo {volume} {79}},\
  \bibinfo {pages} {733} (\bibinfo {year} {2007})},\ \Eprint
  {http://arxiv.org/abs/hep-th/0610102} {arXiv:hep-th/0610102 [hep-th]}
  \BibitemShut {NoStop}%
%%CITATION = HEP-TH/0610102;%%
\bibitem [{\citenamefont {Conlon}\ and\ \citenamefont
  {Quevedo}(2019)}]{Conlon:2018vov}%
  \BibitemOpen
  \bibfield  {author} {\bibinfo {author} {\bibfnamefont {J.~P.}\ \bibnamefont
  {Conlon}}\ and\ \bibinfo {author} {\bibfnamefont {F.}~\bibnamefont
  {Quevedo}},\ }\href {\doibase 10.1007/JHEP03(2019)005} {\bibfield  {journal}
  {\bibinfo  {journal} {JHEP}\ }\textbf {\bibinfo {volume} {03}},\ \bibinfo
  {pages} {005} (\bibinfo {year} {2019})},\ \Eprint
  {http://arxiv.org/abs/1811.06276} {arXiv:1811.06276 [hep-th]} \BibitemShut
  {NoStop}%
%%CITATION = ARXIV:1811.06276;%%
\bibitem [{\citenamefont {Collins}\ \emph {et~al.}()\citenamefont {Collins},
  \citenamefont {Jafferis}, \citenamefont {Vafa},\ and\ \citenamefont
  {Yau}}]{CJVY}%
  \BibitemOpen
  \bibfield  {author} {\bibinfo {author} {\bibfnamefont {T.}~\bibnamefont
  {Collins}}, \bibinfo {author} {\bibfnamefont {D.}~\bibnamefont {Jafferis}},
  \bibinfo {author} {\bibfnamefont {C.}~\bibnamefont {Vafa}}, \ and\ \bibinfo
  {author} {\bibfnamefont {S.~T.}\ \bibnamefont {Yau}},\ }\href@noop {}
  {\bibinfo  {journal} {Work in progress}\ }\BibitemShut {NoStop}%
\bibitem [{\citenamefont {Ooguri}\ and\ \citenamefont
  {Vafa}(2017)}]{Ooguri:2016pdq}%
  \BibitemOpen
\bibfield  {journal} {  }\bibfield  {author} {\bibinfo {author} {\bibfnamefont
  {H.}~\bibnamefont {Ooguri}}\ and\ \bibinfo {author} {\bibfnamefont
  {C.}~\bibnamefont {Vafa}},\ }\href {\doibase 10.4310/ATMP.2017.v21.n7.a8}
  {\bibfield  {journal} {\bibinfo  {journal} {Adv. Theor. Math. Phys.}\
  }\textbf {\bibinfo {volume} {21}},\ \bibinfo {pages} {1787} (\bibinfo {year}
  {2017})},\ \Eprint {http://arxiv.org/abs/1610.01533} {arXiv:1610.01533
  [hep-th]} \BibitemShut {NoStop}%
%%CITATION = ARXIV:1610.01533;%%
\bibitem [{\citenamefont {Ooguri}\ \emph {et~al.}(2019)\citenamefont {Ooguri},
  \citenamefont {Palti}, \citenamefont {Shiu},\ and\ \citenamefont
  {Vafa}}]{Ooguri:2018wrx}%
  \BibitemOpen
  \bibfield  {author} {\bibinfo {author} {\bibfnamefont {H.}~\bibnamefont
  {Ooguri}}, \bibinfo {author} {\bibfnamefont {E.}~\bibnamefont {Palti}},
  \bibinfo {author} {\bibfnamefont {G.}~\bibnamefont {Shiu}}, \ and\ \bibinfo
  {author} {\bibfnamefont {C.}~\bibnamefont {Vafa}},\ }\href {\doibase
  10.1016/j.physletb.2018.11.018} {\bibfield  {journal} {\bibinfo  {journal}
  {Phys. Lett.}\ }\textbf {\bibinfo {volume} {B788}},\ \bibinfo {pages} {180}
  (\bibinfo {year} {2019})},\ \Eprint {http://arxiv.org/abs/1810.05506}
  {arXiv:1810.05506 [hep-th]} \BibitemShut {NoStop}%
%%CITATION = ARXIV:1810.05506;%%
\bibitem [{\citenamefont {Garg}\ and\ \citenamefont
  {Krishnan}(2018)}]{Garg:2018reu}%
  \BibitemOpen
  \bibfield  {author} {\bibinfo {author} {\bibfnamefont {S.~K.}\ \bibnamefont
  {Garg}}\ and\ \bibinfo {author} {\bibfnamefont {C.}~\bibnamefont
  {Krishnan}},\ }\href@noop {} {\  (\bibinfo {year} {2018})},\ \Eprint
  {http://arxiv.org/abs/1807.05193} {arXiv:1807.05193 [hep-th]} \BibitemShut
  {NoStop}%
%%CITATION = ARXIV:1807.05193;%%
\bibitem [{\citenamefont {Arkani-Hamed}\ \emph {et~al.}(2007)\citenamefont
  {Arkani-Hamed}, \citenamefont {Motl}, \citenamefont {Nicolis},\ and\
  \citenamefont {Vafa}}]{ArkaniHamed:2006dz}%
  \BibitemOpen
  \bibfield  {author} {\bibinfo {author} {\bibfnamefont {N.}~\bibnamefont
  {Arkani-Hamed}}, \bibinfo {author} {\bibfnamefont {L.}~\bibnamefont {Motl}},
  \bibinfo {author} {\bibfnamefont {A.}~\bibnamefont {Nicolis}}, \ and\
  \bibinfo {author} {\bibfnamefont {C.}~\bibnamefont {Vafa}},\ }\href {\doibase
  10.1088/1126-6708/2007/06/060} {\bibfield  {journal} {\bibinfo  {journal}
  {JHEP}\ }\textbf {\bibinfo {volume} {06}},\ \bibinfo {pages} {060} (\bibinfo
  {year} {2007})},\ \Eprint {http://arxiv.org/abs/hep-th/0601001}
  {arXiv:hep-th/0601001 [hep-th]} \BibitemShut {NoStop}%
%%CITATION = HEP-TH/0601001;%%
\bibitem [{\citenamefont {Heidenreich}\ \emph {et~al.}(2016)\citenamefont
  {Heidenreich}, \citenamefont {Reece},\ and\ \citenamefont
  {Rudelius}}]{Heidenreich:2015nta}%
  \BibitemOpen
  \bibfield  {author} {\bibinfo {author} {\bibfnamefont {B.}~\bibnamefont
  {Heidenreich}}, \bibinfo {author} {\bibfnamefont {M.}~\bibnamefont {Reece}},
  \ and\ \bibinfo {author} {\bibfnamefont {T.}~\bibnamefont {Rudelius}},\
  }\href {\doibase 10.1007/JHEP02(2016)140} {\bibfield  {journal} {\bibinfo
  {journal} {JHEP}\ }\textbf {\bibinfo {volume} {02}},\ \bibinfo {pages} {140}
  (\bibinfo {year} {2016})},\ \Eprint {http://arxiv.org/abs/1509.06374}
  {arXiv:1509.06374 [hep-th]} \BibitemShut {NoStop}%
%%CITATION = ARXIV:1509.06374;%%
\bibitem [{\citenamefont {Klaewer}\ and\ \citenamefont
  {Palti}(2017)}]{Klaewer:2016kiy}%
  \BibitemOpen
  \bibfield  {author} {\bibinfo {author} {\bibfnamefont {D.}~\bibnamefont
  {Klaewer}}\ and\ \bibinfo {author} {\bibfnamefont {E.}~\bibnamefont
  {Palti}},\ }\href {\doibase 10.1007/JHEP01(2017)088} {\bibfield  {journal}
  {\bibinfo  {journal} {JHEP}\ }\textbf {\bibinfo {volume} {01}},\ \bibinfo
  {pages} {088} (\bibinfo {year} {2017})},\ \Eprint
  {http://arxiv.org/abs/1610.00010} {arXiv:1610.00010 [hep-th]} \BibitemShut
  {NoStop}%
%%CITATION = ARXIV:1610.00010;%%
\bibitem [{\citenamefont {Montero}\ \emph {et~al.}(2016)\citenamefont
  {Montero}, \citenamefont {Shiu},\ and\ \citenamefont
  {Soler}}]{Montero:2016tif}%
  \BibitemOpen
  \bibfield  {author} {\bibinfo {author} {\bibfnamefont {M.}~\bibnamefont
  {Montero}}, \bibinfo {author} {\bibfnamefont {G.}~\bibnamefont {Shiu}}, \
  and\ \bibinfo {author} {\bibfnamefont {P.}~\bibnamefont {Soler}},\ }\href
  {\doibase 10.1007/JHEP10(2016)159} {\bibfield  {journal} {\bibinfo  {journal}
  {JHEP}\ }\textbf {\bibinfo {volume} {10}},\ \bibinfo {pages} {159} (\bibinfo
  {year} {2016})},\ \Eprint {http://arxiv.org/abs/1606.08438} {arXiv:1606.08438
  [hep-th]} \BibitemShut {NoStop}%
%%CITATION = ARXIV:1606.08438;%%
\bibitem [{\citenamefont {Heidenreich}\ \emph {et~al.}(2017)\citenamefont
  {Heidenreich}, \citenamefont {Reece},\ and\ \citenamefont
  {Rudelius}}]{Heidenreich:2016aqi}%
  \BibitemOpen
  \bibfield  {author} {\bibinfo {author} {\bibfnamefont {B.}~\bibnamefont
  {Heidenreich}}, \bibinfo {author} {\bibfnamefont {M.}~\bibnamefont {Reece}},
  \ and\ \bibinfo {author} {\bibfnamefont {T.}~\bibnamefont {Rudelius}},\
  }\href {\doibase 10.1007/JHEP08(2017)025} {\bibfield  {journal} {\bibinfo
  {journal} {JHEP}\ }\textbf {\bibinfo {volume} {08}},\ \bibinfo {pages} {025}
  (\bibinfo {year} {2017})},\ \Eprint {http://arxiv.org/abs/1606.08437}
  {arXiv:1606.08437 [hep-th]} \BibitemShut {NoStop}%
%%CITATION = ARXIV:1606.08437;%%
\bibitem [{\citenamefont {Andriolo}\ \emph {et~al.}(2018)\citenamefont
  {Andriolo}, \citenamefont {Junghans}, \citenamefont {Noumi},\ and\
  \citenamefont {Shiu}}]{Andriolo:2018lvp}%
  \BibitemOpen
  \bibfield  {author} {\bibinfo {author} {\bibfnamefont {S.}~\bibnamefont
  {Andriolo}}, \bibinfo {author} {\bibfnamefont {D.}~\bibnamefont {Junghans}},
  \bibinfo {author} {\bibfnamefont {T.}~\bibnamefont {Noumi}}, \ and\ \bibinfo
  {author} {\bibfnamefont {G.}~\bibnamefont {Shiu}},\ }\href {\doibase
  10.1002/prop.201800020} {\bibfield  {journal} {\bibinfo  {journal} {Fortsch.
  Phys.}\ }\textbf {\bibinfo {volume} {66}},\ \bibinfo {pages} {1800020}
  (\bibinfo {year} {2018})},\ \Eprint {http://arxiv.org/abs/1802.04287}
  {arXiv:1802.04287 [hep-th]} \BibitemShut {NoStop}%
%%CITATION = ARXIV:1802.04287;%%
\bibitem [{\citenamefont {Grimm}\ \emph {et~al.}(2018)\citenamefont {Grimm},
  \citenamefont {Palti},\ and\ \citenamefont {Valenzuela}}]{Grimm:2018ohb}%
  \BibitemOpen
  \bibfield  {author} {\bibinfo {author} {\bibfnamefont {T.~W.}\ \bibnamefont
  {Grimm}}, \bibinfo {author} {\bibfnamefont {E.}~\bibnamefont {Palti}}, \ and\
  \bibinfo {author} {\bibfnamefont {I.}~\bibnamefont {Valenzuela}},\ }\href
  {\doibase 10.1007/JHEP08(2018)143} {\bibfield  {journal} {\bibinfo  {journal}
  {JHEP}\ }\textbf {\bibinfo {volume} {08}},\ \bibinfo {pages} {143} (\bibinfo
  {year} {2018})},\ \Eprint {http://arxiv.org/abs/1802.08264} {arXiv:1802.08264
  [hep-th]} \BibitemShut {NoStop}%
%%CITATION = ARXIV:1802.08264;%%
\bibitem [{\citenamefont {Ooguri}\ and\ \citenamefont
  {Vafa}(2007)}]{Ooguri:2006in}%
  \BibitemOpen
  \bibfield  {author} {\bibinfo {author} {\bibfnamefont {H.}~\bibnamefont
  {Ooguri}}\ and\ \bibinfo {author} {\bibfnamefont {C.}~\bibnamefont {Vafa}},\
  }\href {\doibase 10.1016/j.nuclphysb.2006.10.033} {\bibfield  {journal}
  {\bibinfo  {journal} {Nucl. Phys.}\ }\textbf {\bibinfo {volume} {B766}},\
  \bibinfo {pages} {21} (\bibinfo {year} {2007})},\ \Eprint
  {http://arxiv.org/abs/hep-th/0605264} {arXiv:hep-th/0605264 [hep-th]}
  \BibitemShut {NoStop}%
%%CITATION = HEP-TH/0605264;%%
\bibitem [{\citenamefont {Baume}\ and\ \citenamefont
  {Palti}(2016)}]{Baume:2016psm}%
  \BibitemOpen
  \bibfield  {author} {\bibinfo {author} {\bibfnamefont {F.}~\bibnamefont
  {Baume}}\ and\ \bibinfo {author} {\bibfnamefont {E.}~\bibnamefont {Palti}},\
  }\href {\doibase 10.1007/JHEP08(2016)043} {\bibfield  {journal} {\bibinfo
  {journal} {JHEP}\ }\textbf {\bibinfo {volume} {08}},\ \bibinfo {pages} {043}
  (\bibinfo {year} {2016})},\ \Eprint {http://arxiv.org/abs/1602.06517}
  {arXiv:1602.06517 [hep-th]} \BibitemShut {NoStop}%
%%CITATION = ARXIV:1602.06517;%%
\bibitem [{\citenamefont {Cecotti}(2015)}]{Cecotti:2015wqa}%
  \BibitemOpen
  \bibfield  {author} {\bibinfo {author} {\bibfnamefont {S.}~\bibnamefont
  {Cecotti}},\ }\href@noop {} {\emph {\bibinfo {title} {{Supersymmetric Field
  Theories}}}}\ (\bibinfo  {publisher} {Cambridge University Press},\ \bibinfo
  {year} {2015})\BibitemShut {NoStop}%
%%CITATION = INSPIRE-1384886;%%
\bibitem [{\citenamefont {Palti}(2015)}]{Palti:2015xra}%
  \BibitemOpen
  \bibfield  {author} {\bibinfo {author} {\bibfnamefont {E.}~\bibnamefont
  {Palti}},\ }\href {\doibase 10.1007/JHEP10(2015)188} {\bibfield  {journal}
  {\bibinfo  {journal} {JHEP}\ }\textbf {\bibinfo {volume} {10}},\ \bibinfo
  {pages} {188} (\bibinfo {year} {2015})},\ \Eprint
  {http://arxiv.org/abs/1508.00009} {arXiv:1508.00009 [hep-th]} \BibitemShut
  {NoStop}%
%%CITATION = ARXIV:1508.00009;%%
\bibitem [{\citenamefont
  {Valenzuela}(2017{\natexlab{a}})}]{Valenzuela:2016yny}%
  \BibitemOpen
  \bibfield  {author} {\bibinfo {author} {\bibfnamefont {I.}~\bibnamefont
  {Valenzuela}},\ }\href {\doibase 10.1007/JHEP06(2017)098} {\bibfield
  {journal} {\bibinfo  {journal} {JHEP}\ }\textbf {\bibinfo {volume} {06}},\
  \bibinfo {pages} {098} (\bibinfo {year} {2017}{\natexlab{a}})},\ \Eprint
  {http://arxiv.org/abs/1611.00394} {arXiv:1611.00394 [hep-th]} \BibitemShut
  {NoStop}%
%%CITATION = ARXIV:1611.00394;%%
\bibitem [{\citenamefont {Blumenhagen}\ \emph {et~al.}(2017)\citenamefont
  {Blumenhagen}, \citenamefont {Valenzuela},\ and\ \citenamefont
  {Wolf}}]{Blumenhagen:2017cxt}%
  \BibitemOpen
  \bibfield  {author} {\bibinfo {author} {\bibfnamefont {R.}~\bibnamefont
  {Blumenhagen}}, \bibinfo {author} {\bibfnamefont {I.}~\bibnamefont
  {Valenzuela}}, \ and\ \bibinfo {author} {\bibfnamefont {F.}~\bibnamefont
  {Wolf}},\ }\href {\doibase 10.1007/JHEP07(2017)145} {\bibfield  {journal}
  {\bibinfo  {journal} {JHEP}\ }\textbf {\bibinfo {volume} {07}},\ \bibinfo
  {pages} {145} (\bibinfo {year} {2017})},\ \Eprint
  {http://arxiv.org/abs/1703.05776} {arXiv:1703.05776 [hep-th]} \BibitemShut
  {NoStop}%
%%CITATION = ARXIV:1703.05776;%%
\bibitem [{\citenamefont
  {Valenzuela}(2017{\natexlab{b}})}]{Valenzuela:2017bvg}%
  \BibitemOpen
  \bibfield  {author} {\bibinfo {author} {\bibfnamefont {I.}~\bibnamefont
  {Valenzuela}},\ }\bibfield  {booktitle} {\emph {\bibinfo {booktitle}
  {{Proceedings, 16th Hellenic School and Workshops on Elementary Particle
  Physics and Gravity (CORFU2016): Corfu, Greece, August 31-September 23,
  2016}}},\ }\href {\doibase 10.22323/1.292.0112} {\bibfield  {journal}
  {\bibinfo  {journal} {PoS}\ }\textbf {\bibinfo {volume} {CORFU2016}},\
  \bibinfo {pages} {112} (\bibinfo {year} {2017}{\natexlab{b}})},\ \Eprint
  {http://arxiv.org/abs/1708.07456} {arXiv:1708.07456 [hep-th]} \BibitemShut
  {NoStop}%
%%CITATION = ARXIV:1708.07456;%%
\bibitem [{\citenamefont {Palti}(2017)}]{Palti:2017elp}%
  \BibitemOpen
  \bibfield  {author} {\bibinfo {author} {\bibfnamefont {E.}~\bibnamefont
  {Palti}},\ }\href {\doibase 10.1007/JHEP08(2017)034} {\bibfield  {journal}
  {\bibinfo  {journal} {JHEP}\ }\textbf {\bibinfo {volume} {08}},\ \bibinfo
  {pages} {034} (\bibinfo {year} {2017})},\ \Eprint
  {http://arxiv.org/abs/1705.04328} {arXiv:1705.04328 [hep-th]} \BibitemShut
  {NoStop}%
%%CITATION = ARXIV:1705.04328;%%
\bibitem [{\citenamefont {Hebecker}\ \emph {et~al.}(2017)\citenamefont
  {Hebecker}, \citenamefont {Henkenjohann},\ and\ \citenamefont
  {Witkowski}}]{Hebecker:2017lxm}%
  \BibitemOpen
  \bibfield  {author} {\bibinfo {author} {\bibfnamefont {A.}~\bibnamefont
  {Hebecker}}, \bibinfo {author} {\bibfnamefont {P.}~\bibnamefont
  {Henkenjohann}}, \ and\ \bibinfo {author} {\bibfnamefont {L.~T.}\
  \bibnamefont {Witkowski}},\ }\href {\doibase 10.1007/JHEP12(2017)033}
  {\bibfield  {journal} {\bibinfo  {journal} {JHEP}\ }\textbf {\bibinfo
  {volume} {12}},\ \bibinfo {pages} {033} (\bibinfo {year} {2017})},\ \Eprint
  {http://arxiv.org/abs/1708.06761} {arXiv:1708.06761 [hep-th]} \BibitemShut
  {NoStop}%
%%CITATION = ARXIV:1708.06761;%%
\bibitem [{\citenamefont {Cicoli}\ \emph {et~al.}(2018)\citenamefont {Cicoli},
  \citenamefont {Ciupke}, \citenamefont {Mayrhofer},\ and\ \citenamefont
  {Shukla}}]{Cicoli:2018tcq}%
  \BibitemOpen
  \bibfield  {author} {\bibinfo {author} {\bibfnamefont {M.}~\bibnamefont
  {Cicoli}}, \bibinfo {author} {\bibfnamefont {D.}~\bibnamefont {Ciupke}},
  \bibinfo {author} {\bibfnamefont {C.}~\bibnamefont {Mayrhofer}}, \ and\
  \bibinfo {author} {\bibfnamefont {P.}~\bibnamefont {Shukla}},\ }\href
  {\doibase 10.1007/JHEP05(2018)001} {\bibfield  {journal} {\bibinfo  {journal}
  {JHEP}\ }\textbf {\bibinfo {volume} {05}},\ \bibinfo {pages} {001} (\bibinfo
  {year} {2018})},\ \Eprint {http://arxiv.org/abs/1801.05434} {arXiv:1801.05434
  [hep-th]} \BibitemShut {NoStop}%
%%CITATION = ARXIV:1801.05434;%%
\bibitem [{\citenamefont {Blumenhagen}\ \emph {et~al.}(2018)\citenamefont
  {Blumenhagen}, \citenamefont {Kl{\"a}wer}, \citenamefont {Schlechter},\ and\
  \citenamefont {Wolf}}]{Blumenhagen:2018nts}%
  \BibitemOpen
  \bibfield  {author} {\bibinfo {author} {\bibfnamefont {R.}~\bibnamefont
  {Blumenhagen}}, \bibinfo {author} {\bibfnamefont {D.}~\bibnamefont
  {Kl{\"a}wer}}, \bibinfo {author} {\bibfnamefont {L.}~\bibnamefont
  {Schlechter}}, \ and\ \bibinfo {author} {\bibfnamefont {F.}~\bibnamefont
  {Wolf}},\ }\href {\doibase 10.1007/JHEP06(2018)052} {\bibfield  {journal}
  {\bibinfo  {journal} {JHEP}\ }\textbf {\bibinfo {volume} {06}},\ \bibinfo
  {pages} {052} (\bibinfo {year} {2018})},\ \Eprint
  {http://arxiv.org/abs/1803.04989} {arXiv:1803.04989 [hep-th]} \BibitemShut
  {NoStop}%
%%CITATION = ARXIV:1803.04989;%%
\bibitem [{\citenamefont {Blumenhagen}(2018)}]{Blumenhagen:2018hsh}%
  \BibitemOpen
  \bibfield  {author} {\bibinfo {author} {\bibfnamefont {R.}~\bibnamefont
  {Blumenhagen}},\ }\bibfield  {booktitle} {\emph {\bibinfo {booktitle}
  {{Proceedings, 17th Hellenic School and Workshops on Elementary Particle
  Physics and Gravity (CORFU2017): Corfu, Greece, September 2-28, 2017}}},\
  }\href {\doibase 10.22323/1.318.0175} {\bibfield  {journal} {\bibinfo
  {journal} {PoS}\ }\textbf {\bibinfo {volume} {CORFU2017}},\ \bibinfo {pages}
  {175} (\bibinfo {year} {2018})},\ \Eprint {http://arxiv.org/abs/1804.10504}
  {arXiv:1804.10504 [hep-th]} \BibitemShut {NoStop}%
%%CITATION = ARXIV:1804.10504;%%
\bibitem [{\citenamefont {Lee}\ \emph {et~al.}(2018)\citenamefont {Lee},
  \citenamefont {Lerche},\ and\ \citenamefont {Weigand}}]{Lee:2018urn}%
  \BibitemOpen
  \bibfield  {author} {\bibinfo {author} {\bibfnamefont {S.-J.}\ \bibnamefont
  {Lee}}, \bibinfo {author} {\bibfnamefont {W.}~\bibnamefont {Lerche}}, \ and\
  \bibinfo {author} {\bibfnamefont {T.}~\bibnamefont {Weigand}},\ }\href
  {\doibase 10.1007/JHEP10(2018)164} {\bibfield  {journal} {\bibinfo  {journal}
  {JHEP}\ }\textbf {\bibinfo {volume} {10}},\ \bibinfo {pages} {164} (\bibinfo
  {year} {2018})},\ \Eprint {http://arxiv.org/abs/1808.05958} {arXiv:1808.05958
  [hep-th]} \BibitemShut {NoStop}%
%%CITATION = ARXIV:1808.05958;%%
\bibitem [{\citenamefont {Lee}\ \emph {et~al.}(2019{\natexlab{a}})\citenamefont
  {Lee}, \citenamefont {Lerche},\ and\ \citenamefont {Weigand}}]{Lee:2018spm}%
  \BibitemOpen
  \bibfield  {author} {\bibinfo {author} {\bibfnamefont {S.-J.}\ \bibnamefont
  {Lee}}, \bibinfo {author} {\bibfnamefont {W.}~\bibnamefont {Lerche}}, \ and\
  \bibinfo {author} {\bibfnamefont {T.}~\bibnamefont {Weigand}},\ }\href
  {\doibase 10.1016/j.nuclphysb.2018.11.001} {\bibfield  {journal} {\bibinfo
  {journal} {Nucl. Phys.}\ }\textbf {\bibinfo {volume} {B938}},\ \bibinfo
  {pages} {321} (\bibinfo {year} {2019}{\natexlab{a}})},\ \Eprint
  {http://arxiv.org/abs/1810.05169} {arXiv:1810.05169 [hep-th]} \BibitemShut
  {NoStop}%
%%CITATION = ARXIV:1810.05169;%%
\bibitem [{\citenamefont {Grimm}\ \emph {et~al.}(2019)\citenamefont {Grimm},
  \citenamefont {Li},\ and\ \citenamefont {Palti}}]{Grimm:2018cpv}%
  \BibitemOpen
  \bibfield  {author} {\bibinfo {author} {\bibfnamefont {T.~W.}\ \bibnamefont
  {Grimm}}, \bibinfo {author} {\bibfnamefont {C.}~\bibnamefont {Li}}, \ and\
  \bibinfo {author} {\bibfnamefont {E.}~\bibnamefont {Palti}},\ }\href
  {\doibase 10.1007/JHEP03(2019)016} {\bibfield  {journal} {\bibinfo  {journal}
  {JHEP}\ }\textbf {\bibinfo {volume} {03}},\ \bibinfo {pages} {016} (\bibinfo
  {year} {2019})},\ \Eprint {http://arxiv.org/abs/1811.02571} {arXiv:1811.02571
  [hep-th]} \BibitemShut {NoStop}%
%%CITATION = ARXIV:1811.02571;%%
\bibitem [{\citenamefont {Kim}\ and\ \citenamefont
  {McAllister}(2018)}]{Kim:2018vgz}%
  \BibitemOpen
  \bibfield  {author} {\bibinfo {author} {\bibfnamefont {M.}~\bibnamefont
  {Kim}}\ and\ \bibinfo {author} {\bibfnamefont {L.}~\bibnamefont
  {McAllister}},\ }\href@noop {} {\  (\bibinfo {year} {2018})},\ \Eprint
  {http://arxiv.org/abs/1812.03532} {arXiv:1812.03532 [hep-th]} \BibitemShut
  {NoStop}%
%%CITATION = ARXIV:1812.03532;%%
\bibitem [{\citenamefont {Buratti}\ \emph {et~al.}(2018)\citenamefont
  {Buratti}, \citenamefont {Calderón},\ and\ \citenamefont
  {Uranga}}]{Buratti:2018xjt}%
  \BibitemOpen
  \bibfield  {author} {\bibinfo {author} {\bibfnamefont {G.}~\bibnamefont
  {Buratti}}, \bibinfo {author} {\bibfnamefont {J.}~\bibnamefont {Calderón}}, \
  and\ \bibinfo {author} {\bibfnamefont {A.~M.}\ \bibnamefont {Uranga}},\
  }\href@noop {} {\  (\bibinfo {year} {2018})},\ \Eprint
  {http://arxiv.org/abs/1812.05016} {arXiv:1812.05016 [hep-th]} \BibitemShut
  {NoStop}%
%%CITATION = ARXIV:1812.05016;%%
\bibitem [{\citenamefont {Hebecker}\ \emph {et~al.}(2019)\citenamefont
  {Hebecker}, \citenamefont {Junghans},\ and\ \citenamefont
  {Schachner}}]{Hebecker:2018fln}%
  \BibitemOpen
  \bibfield  {author} {\bibinfo {author} {\bibfnamefont {A.}~\bibnamefont
  {Hebecker}}, \bibinfo {author} {\bibfnamefont {D.}~\bibnamefont {Junghans}},
  \ and\ \bibinfo {author} {\bibfnamefont {A.}~\bibnamefont {Schachner}},\
  }\href {\doibase 10.1007/JHEP03(2019)192} {\bibfield  {journal} {\bibinfo
  {journal} {JHEP}\ }\textbf {\bibinfo {volume} {03}},\ \bibinfo {pages} {192}
  (\bibinfo {year} {2019})},\ \Eprint {http://arxiv.org/abs/1812.05626}
  {arXiv:1812.05626 [hep-th]} \BibitemShut {NoStop}%
%%CITATION = ARXIV:1812.05626;%%
\bibitem [{\citenamefont {Corvilain}\ \emph {et~al.}(2018)\citenamefont
  {Corvilain}, \citenamefont {Grimm},\ and\ \citenamefont
  {Valenzuela}}]{Corvilain:2018lgw}%
  \BibitemOpen
  \bibfield  {author} {\bibinfo {author} {\bibfnamefont {P.}~\bibnamefont
  {Corvilain}}, \bibinfo {author} {\bibfnamefont {T.~W.}\ \bibnamefont
  {Grimm}}, \ and\ \bibinfo {author} {\bibfnamefont {I.}~\bibnamefont
  {Valenzuela}},\ }\href@noop {} {\  (\bibinfo {year} {2018})},\ \Eprint
  {http://arxiv.org/abs/1812.07548} {arXiv:1812.07548 [hep-th]} \BibitemShut
  {NoStop}%
%%CITATION = ARXIV:1812.07548;%%
\bibitem [{\citenamefont {Joshi}\ and\ \citenamefont
  {Klemm}(2019)}]{Joshi:2019nzi}%
  \BibitemOpen
  \bibfield  {author} {\bibinfo {author} {\bibfnamefont {A.}~\bibnamefont
  {Joshi}}\ and\ \bibinfo {author} {\bibfnamefont {A.}~\bibnamefont {Klemm}},\
  }\href@noop {} {\  (\bibinfo {year} {2019})},\ \Eprint
  {http://arxiv.org/abs/1903.00596} {arXiv:1903.00596 [hep-th]} \BibitemShut
  {NoStop}%
%%CITATION = ARXIV:1903.00596;%%
\bibitem [{\citenamefont {Scalisi}\ and\ \citenamefont
  {Valenzuela}(2018)}]{Scalisi:2018eaz}%
  \BibitemOpen
  \bibfield  {author} {\bibinfo {author} {\bibfnamefont {M.}~\bibnamefont
  {Scalisi}}\ and\ \bibinfo {author} {\bibfnamefont {I.}~\bibnamefont
  {Valenzuela}},\ }\href@noop {} {\  (\bibinfo {year} {2018})},\ \Eprint
  {http://arxiv.org/abs/1812.07558} {arXiv:1812.07558 [hep-th]} \BibitemShut
  {NoStop}%
%%CITATION = ARXIV:1812.07558;%%
\bibitem [{\citenamefont {Lee}\ \emph {et~al.}(2019{\natexlab{b}})\citenamefont
  {Lee}, \citenamefont {Lerche},\ and\ \citenamefont {Weigand}}]{Lee:2019tst}%
  \BibitemOpen
  \bibfield  {author} {\bibinfo {author} {\bibfnamefont {S.-J.}\ \bibnamefont
  {Lee}}, \bibinfo {author} {\bibfnamefont {W.}~\bibnamefont {Lerche}}, \ and\
  \bibinfo {author} {\bibfnamefont {T.}~\bibnamefont {Weigand}},\ }\href@noop
  {} {\  (\bibinfo {year} {2019}{\natexlab{b}})},\ \Eprint
  {http://arxiv.org/abs/1901.08065} {arXiv:1901.08065 [hep-th]} \BibitemShut
  {NoStop}%
%%CITATION = ARXIV:1901.08065;%%
\bibitem [{\citenamefont {Blumenhagen}\ \emph {et~al.}(2019)\citenamefont
  {Blumenhagen}, \citenamefont {Kl{\"a}wer},\ and\ \citenamefont
  {Schlechter}}]{Blumenhagen:2019qcg}%
  \BibitemOpen
  \bibfield  {author} {\bibinfo {author} {\bibfnamefont {R.}~\bibnamefont
  {Blumenhagen}}, \bibinfo {author} {\bibfnamefont {D.}~\bibnamefont
  {Kl{\"a}wer}}, \ and\ \bibinfo {author} {\bibfnamefont {L.}~\bibnamefont
  {Schlechter}},\ }\href {\doibase 10.1007/JHEP05(2019)152} {\bibfield
  {journal} {\bibinfo  {journal} {JHEP}\ }\textbf {\bibinfo {volume} {05}},\
  \bibinfo {pages} {152} (\bibinfo {year} {2019})},\ \Eprint
  {http://arxiv.org/abs/1902.07724} {arXiv:1902.07724 [hep-th]} \BibitemShut
  {NoStop}%
%%CITATION = ARXIV:1902.07724;%%
\bibitem [{\citenamefont {Marchesano}\ and\ \citenamefont
  {Wiesner}(2019)}]{Marchesano:2019ifh}%
  \BibitemOpen
  \bibfield  {author} {\bibinfo {author} {\bibfnamefont {F.}~\bibnamefont
  {Marchesano}}\ and\ \bibinfo {author} {\bibfnamefont {M.}~\bibnamefont
  {Wiesner}},\ }\href@noop {} {\  (\bibinfo {year} {2019})},\ \Eprint
  {http://arxiv.org/abs/1904.04848} {arXiv:1904.04848 [hep-th]} \BibitemShut
  {NoStop}%
%%CITATION = ARXIV:1904.04848;%%
\bibitem [{\citenamefont {Font}\ \emph {et~al.}(2019)\citenamefont {Font},
  \citenamefont {Herráez},\ and\ \citenamefont {Ibanez}}]{Font:2019cxq}%
  \BibitemOpen
  \bibfield  {author} {\bibinfo {author} {\bibfnamefont {A.}~\bibnamefont
  {Font}}, \bibinfo {author} {\bibfnamefont {A.}~\bibnamefont {Herráez}}, \
  and\ \bibinfo {author} {\bibfnamefont {L.~E.}\ \bibnamefont {Ibanez}},\
  }\href@noop {} {\  (\bibinfo {year} {2019})},\ \Eprint
  {http://arxiv.org/abs/1904.05379} {arXiv:1904.05379 [hep-th]} \BibitemShut
  {NoStop}%
%%CITATION = ARXIV:1904.05379;%%
\bibitem [{\citenamefont {Lee}\ \emph {et~al.}(2019{\natexlab{c}})\citenamefont
  {Lee}, \citenamefont {Lerche},\ and\ \citenamefont {Weigand}}]{Lee:2019xtm}%
  \BibitemOpen
  \bibfield  {author} {\bibinfo {author} {\bibfnamefont {S.-J.}\ \bibnamefont
  {Lee}}, \bibinfo {author} {\bibfnamefont {W.}~\bibnamefont {Lerche}}, \ and\
  \bibinfo {author} {\bibfnamefont {T.}~\bibnamefont {Weigand}},\ }\href@noop
  {} {\  (\bibinfo {year} {2019}{\natexlab{c}})},\ \Eprint
  {http://arxiv.org/abs/1904.06344} {arXiv:1904.06344 [hep-th]} \BibitemShut
  {NoStop}%
%%CITATION = ARXIV:1904.06344;%%
\bibitem [{\citenamefont {Grimm}\ and\ \citenamefont {Van
  De~Heisteeg}(2019)}]{Grimm:2019wtx}%
  \BibitemOpen
  \bibfield  {author} {\bibinfo {author} {\bibfnamefont {T.~W.}\ \bibnamefont
  {Grimm}}\ and\ \bibinfo {author} {\bibfnamefont {D.}~\bibnamefont {Van
  De~Heisteeg}},\ }\href@noop {} {\  (\bibinfo {year} {2019})},\ \Eprint
  {http://arxiv.org/abs/1905.00901} {arXiv:1905.00901 [hep-th]} \BibitemShut
  {NoStop}%
%%CITATION = ARXIV:1905.00901;%%
\bibitem [{\citenamefont {Erkinger}\ and\ \citenamefont
  {Knapp}(2019)}]{Erkinger:2019umg}%
  \BibitemOpen
  \bibfield  {author} {\bibinfo {author} {\bibfnamefont {D.}~\bibnamefont
  {Erkinger}}\ and\ \bibinfo {author} {\bibfnamefont {J.}~\bibnamefont
  {Knapp}},\ }\href@noop {} {\  (\bibinfo {year} {2019})},\ \Eprint
  {http://arxiv.org/abs/1905.05225} {arXiv:1905.05225 [hep-th]} \BibitemShut
  {NoStop}%
%%CITATION = ARXIV:1905.05225;%%
\bibitem [{\citenamefont {Heidenreich}\ \emph {et~al.}(2018)\citenamefont
  {Heidenreich}, \citenamefont {Reece},\ and\ \citenamefont
  {Rudelius}}]{Heidenreich:2018kpg}%
  \BibitemOpen
  \bibfield  {author} {\bibinfo {author} {\bibfnamefont {B.}~\bibnamefont
  {Heidenreich}}, \bibinfo {author} {\bibfnamefont {M.}~\bibnamefont {Reece}},
  \ and\ \bibinfo {author} {\bibfnamefont {T.}~\bibnamefont {Rudelius}},\
  }\href {\doibase 10.1103/PhysRevLett.121.051601} {\bibfield  {journal}
  {\bibinfo  {journal} {Phys. Rev. Lett.}\ }\textbf {\bibinfo {volume} {121}},\
  \bibinfo {pages} {051601} (\bibinfo {year} {2018})},\ \Eprint
  {http://arxiv.org/abs/1802.08698} {arXiv:1802.08698 [hep-th]} \BibitemShut
  {NoStop}%
%%CITATION = ARXIV:1802.08698;%%
\bibitem [{\citenamefont {DeWitt}(1967)}]{PhysRev.160.1113}%
  \BibitemOpen
  \bibfield  {author} {\bibinfo {author} {\bibfnamefont {B.~S.}\ \bibnamefont
  {DeWitt}},\ }\href {\doibase 10.1103/PhysRev.160.1113} {\bibfield  {journal}
  {\bibinfo  {journal} {Phys. Rev.}\ }\textbf {\bibinfo {volume} {160}},\
  \bibinfo {pages} {1113} (\bibinfo {year} {1967})}\BibitemShut {NoStop}%
\bibitem [{\citenamefont {{Gil-Medrano}}\ and\ \citenamefont
  {{Michor}}(1991)}]{1992math......1259G}%
  \BibitemOpen
  \bibfield  {author} {\bibinfo {author} {\bibfnamefont {O.}~\bibnamefont
  {{Gil-Medrano}}}\ and\ \bibinfo {author} {\bibfnamefont {P.~W.}\ \bibnamefont
  {{Michor}}},\ }\href@noop {} {\bibfield  {journal} {\bibinfo  {journal}
  {arXiv Mathematics e-prints}\ ,\ \bibinfo {eid} {math/9201259}} (\bibinfo
  {year} {1991})},\ \Eprint {http://arxiv.org/abs/math/9201259}
  {arXiv:math/9201259 [math.DG]} \BibitemShut {NoStop}%
\bibitem [{\citenamefont {Candelas}\ and\ \citenamefont {de~la
  Ossa}(1991)}]{Candelas:1990pi}%
  \BibitemOpen
  \bibfield  {author} {\bibinfo {author} {\bibfnamefont {P.}~\bibnamefont
  {Candelas}}\ and\ \bibinfo {author} {\bibfnamefont {X.}~\bibnamefont {de~la
  Ossa}},\ }\href {\doibase 10.1016/0550-3213(91)90122-E} {\bibfield  {journal}
  {\bibinfo  {journal} {Nucl. Phys.}\ }\textbf {\bibinfo {volume} {B355}},\
  \bibinfo {pages} {455} (\bibinfo {year} {1991})}\BibitemShut {NoStop}%
%%CITATION = NUPHA,B355,455;%%
\bibitem [{\citenamefont {Acharya}\ and\ \citenamefont
  {Douglas}(2006)}]{Acharya:2006zw}%
  \BibitemOpen
  \bibfield  {author} {\bibinfo {author} {\bibfnamefont {B.~S.}\ \bibnamefont
  {Acharya}}\ and\ \bibinfo {author} {\bibfnamefont {M.~R.}\ \bibnamefont
  {Douglas}},\ }\href@noop {} {\  (\bibinfo {year} {2006})},\ \Eprint
  {http://arxiv.org/abs/hep-th/0606212} {arXiv:hep-th/0606212 [hep-th]}
  \BibitemShut {NoStop}%
%%CITATION = HEP-TH/0606212;%%
\bibitem [{\citenamefont {Heckman}\ and\ \citenamefont
  {Vafa}(2019)}]{Heckman:2019bzm}%
  \BibitemOpen
  \bibfield  {author} {\bibinfo {author} {\bibfnamefont {J.~J.}\ \bibnamefont
  {Heckman}}\ and\ \bibinfo {author} {\bibfnamefont {C.}~\bibnamefont {Vafa}},\
  }\href@noop {} {\  (\bibinfo {year} {2019})},\ \Eprint
  {http://arxiv.org/abs/1905.06342} {arXiv:1905.06342 [hep-th]} \BibitemShut
  {NoStop}%
%%CITATION = ARXIV:1905.06342;%%
\bibitem [{\citenamefont {Hanany}\ and\ \citenamefont
  {Zaffaroni}(1998{\natexlab{a}})}]{Hanany:1997sa}%
  \BibitemOpen
  \bibfield  {author} {\bibinfo {author} {\bibfnamefont {A.}~\bibnamefont
  {Hanany}}\ and\ \bibinfo {author} {\bibfnamefont {A.}~\bibnamefont
  {Zaffaroni}},\ }\href {\doibase 10.1016/S0550-3213(97)00595-6} {\bibfield
  {journal} {\bibinfo  {journal} {Nucl. Phys.}\ }\textbf {\bibinfo {volume}
  {B509}},\ \bibinfo {pages} {145} (\bibinfo {year} {1998}{\natexlab{a}})},\
  \Eprint {http://arxiv.org/abs/hep-th/9706047} {arXiv:hep-th/9706047 [hep-th]}
  \BibitemShut {NoStop}%
%%CITATION = HEP-TH/9706047;%%
\bibitem [{\citenamefont {Brunner}\ and\ \citenamefont
  {Karch}(1998)}]{Brunner:1997gf}%
  \BibitemOpen
  \bibfield  {author} {\bibinfo {author} {\bibfnamefont {I.}~\bibnamefont
  {Brunner}}\ and\ \bibinfo {author} {\bibfnamefont {A.}~\bibnamefont
  {Karch}},\ }\href {\doibase 10.1088/1126-6708/1998/03/003} {\bibfield
  {journal} {\bibinfo  {journal} {JHEP}\ }\textbf {\bibinfo {volume} {03}},\
  \bibinfo {pages} {003} (\bibinfo {year} {1998})},\ \Eprint
  {http://arxiv.org/abs/hep-th/9712143} {arXiv:hep-th/9712143 [hep-th]}
  \BibitemShut {NoStop}%
%%CITATION = HEP-TH/9712143;%%
\bibitem [{\citenamefont {Ahn}\ \emph {et~al.}(1998)\citenamefont {Ahn},
  \citenamefont {Oh},\ and\ \citenamefont {Tatar}}]{Ahn:1998pb}%
  \BibitemOpen
  \bibfield  {author} {\bibinfo {author} {\bibfnamefont {C.-h.}\ \bibnamefont
  {Ahn}}, \bibinfo {author} {\bibfnamefont {K.}~\bibnamefont {Oh}}, \ and\
  \bibinfo {author} {\bibfnamefont {R.}~\bibnamefont {Tatar}},\ }\href
  {\doibase 10.1016/S0370-2693(98)01276-3} {\bibfield  {journal} {\bibinfo
  {journal} {Phys. Lett.}\ }\textbf {\bibinfo {volume} {B442}},\ \bibinfo
  {pages} {109} (\bibinfo {year} {1998})},\ \Eprint
  {http://arxiv.org/abs/hep-th/9804093} {arXiv:hep-th/9804093 [hep-th]}
  \BibitemShut {NoStop}%
%%CITATION = HEP-TH/9804093;%%
\bibitem [{\citenamefont {Hanany}\ and\ \citenamefont
  {Zaffaroni}(1998{\natexlab{b}})}]{Hanany:1997gh}%
  \BibitemOpen
  \bibfield  {author} {\bibinfo {author} {\bibfnamefont {A.}~\bibnamefont
  {Hanany}}\ and\ \bibinfo {author} {\bibfnamefont {A.}~\bibnamefont
  {Zaffaroni}},\ }\href {\doibase 10.1016/S0550-3213(98)00355-1} {\bibfield
  {journal} {\bibinfo  {journal} {Nucl. Phys.}\ }\textbf {\bibinfo {volume}
  {B529}},\ \bibinfo {pages} {180} (\bibinfo {year} {1998}{\natexlab{b}})},\
  \Eprint {http://arxiv.org/abs/hep-th/9712145} {arXiv:hep-th/9712145 [hep-th]}
  \BibitemShut {NoStop}%
%%CITATION = HEP-TH/9712145;%%
\bibitem [{\citenamefont {Ferrara}\ \emph {et~al.}(1998)\citenamefont
  {Ferrara}, \citenamefont {Kehagias}, \citenamefont {Partouche},\ and\
  \citenamefont {Zaffaroni}}]{Ferrara:1998vf}%
  \BibitemOpen
  \bibfield  {author} {\bibinfo {author} {\bibfnamefont {S.}~\bibnamefont
  {Ferrara}}, \bibinfo {author} {\bibfnamefont {A.}~\bibnamefont {Kehagias}},
  \bibinfo {author} {\bibfnamefont {H.}~\bibnamefont {Partouche}}, \ and\
  \bibinfo {author} {\bibfnamefont {A.}~\bibnamefont {Zaffaroni}},\ }\href
  {\doibase 10.1016/S0370-2693(98)00558-9} {\bibfield  {journal} {\bibinfo
  {journal} {Phys. Lett.}\ }\textbf {\bibinfo {volume} {B431}},\ \bibinfo
  {pages} {42} (\bibinfo {year} {1998})},\ \Eprint
  {http://arxiv.org/abs/hep-th/9803109} {arXiv:hep-th/9803109 [hep-th]}
  \BibitemShut {NoStop}%
%%CITATION = HEP-TH/9803109;%%
\bibitem [{\citenamefont {Apruzzi}\ \emph {et~al.}(2014)\citenamefont
  {Apruzzi}, \citenamefont {Fazzi}, \citenamefont {Rosa},\ and\ \citenamefont
  {Tomasiello}}]{Apruzzi:2013yva}%
  \BibitemOpen
  \bibfield  {author} {\bibinfo {author} {\bibfnamefont {F.}~\bibnamefont
  {Apruzzi}}, \bibinfo {author} {\bibfnamefont {M.}~\bibnamefont {Fazzi}},
  \bibinfo {author} {\bibfnamefont {D.}~\bibnamefont {Rosa}}, \ and\ \bibinfo
  {author} {\bibfnamefont {A.}~\bibnamefont {Tomasiello}},\ }\href {\doibase
  10.1007/JHEP04(2014)064} {\bibfield  {journal} {\bibinfo  {journal} {JHEP}\
  }\textbf {\bibinfo {volume} {04}},\ \bibinfo {pages} {064} (\bibinfo {year}
  {2014})},\ \Eprint {http://arxiv.org/abs/1309.2949} {arXiv:1309.2949
  [hep-th]} \BibitemShut {NoStop}%
%%CITATION = ARXIV:1309.2949;%%
\bibitem [{\citenamefont {Gaiotto}\ and\ \citenamefont
  {Tomasiello}(2014)}]{Gaiotto:2014lca}%
  \BibitemOpen
  \bibfield  {author} {\bibinfo {author} {\bibfnamefont {D.}~\bibnamefont
  {Gaiotto}}\ and\ \bibinfo {author} {\bibfnamefont {A.}~\bibnamefont
  {Tomasiello}},\ }\href {\doibase 10.1007/JHEP12(2014)003} {\bibfield
  {journal} {\bibinfo  {journal} {JHEP}\ }\textbf {\bibinfo {volume} {12}},\
  \bibinfo {pages} {003} (\bibinfo {year} {2014})},\ \Eprint
  {http://arxiv.org/abs/1404.0711} {arXiv:1404.0711 [hep-th]} \BibitemShut
  {NoStop}%
%%CITATION = ARXIV:1404.0711;%%
\bibitem [{\citenamefont {Del~Zotto}\ \emph {et~al.}(2015)\citenamefont
  {Del~Zotto}, \citenamefont {Heckman}, \citenamefont {Tomasiello},\ and\
  \citenamefont {Vafa}}]{DelZotto:2014hpa}%
  \BibitemOpen
  \bibfield  {author} {\bibinfo {author} {\bibfnamefont {M.}~\bibnamefont
  {Del~Zotto}}, \bibinfo {author} {\bibfnamefont {J.~J.}\ \bibnamefont
  {Heckman}}, \bibinfo {author} {\bibfnamefont {A.}~\bibnamefont {Tomasiello}},
  \ and\ \bibinfo {author} {\bibfnamefont {C.}~\bibnamefont {Vafa}},\ }\href
  {\doibase 10.1007/JHEP02(2015)054} {\bibfield  {journal} {\bibinfo  {journal}
  {JHEP}\ }\textbf {\bibinfo {volume} {02}},\ \bibinfo {pages} {054} (\bibinfo
  {year} {2015})},\ \Eprint {http://arxiv.org/abs/1407.6359} {arXiv:1407.6359
  [hep-th]} \BibitemShut {NoStop}%
%%CITATION = ARXIV:1407.6359;%%
\bibitem [{\citenamefont {Louis}\ and\ \citenamefont
  {L{\"u}st}(2015)}]{Louis:2015mka}%
  \BibitemOpen
  \bibfield  {author} {\bibinfo {author} {\bibfnamefont {J.}~\bibnamefont
  {Louis}}\ and\ \bibinfo {author} {\bibfnamefont {S.}~\bibnamefont
  {L{\"u}st}},\ }\href {\doibase 10.1007/JHEP10(2015)120} {\bibfield  {journal}
  {\bibinfo  {journal} {JHEP}\ }\textbf {\bibinfo {volume} {10}},\ \bibinfo
  {pages} {120} (\bibinfo {year} {2015})},\ \Eprint
  {http://arxiv.org/abs/1506.08040} {arXiv:1506.08040 [hep-th]} \BibitemShut
  {NoStop}%
%%CITATION = ARXIV:1506.08040;%%
\bibitem [{\citenamefont {Banks}\ and\ \citenamefont
  {Fischler}(2016)}]{Banks:2016xsj}%
  \BibitemOpen
  \bibfield  {author} {\bibinfo {author} {\bibfnamefont {T.}~\bibnamefont
  {Banks}}\ and\ \bibinfo {author} {\bibfnamefont {W.}~\bibnamefont
  {Fischler}},\ }\href@noop {} {\  (\bibinfo {year} {2016})},\ \Eprint
  {http://arxiv.org/abs/1607.03510} {arXiv:1607.03510 [hep-th]} \BibitemShut
  {NoStop}%
%%CITATION = ARXIV:1607.03510;%%
\bibitem [{\citenamefont {DeWolfe}\ \emph {et~al.}(2005)\citenamefont
  {DeWolfe}, \citenamefont {Giryavets}, \citenamefont {Kachru},\ and\
  \citenamefont {Taylor}}]{DeWolfe:2005uu}%
  \BibitemOpen
  \bibfield  {author} {\bibinfo {author} {\bibfnamefont {O.}~\bibnamefont
  {DeWolfe}}, \bibinfo {author} {\bibfnamefont {A.}~\bibnamefont {Giryavets}},
  \bibinfo {author} {\bibfnamefont {S.}~\bibnamefont {Kachru}}, \ and\ \bibinfo
  {author} {\bibfnamefont {W.}~\bibnamefont {Taylor}},\ }\href {\doibase
  10.1088/1126-6708/2005/07/066} {\bibfield  {journal} {\bibinfo  {journal}
  {JHEP}\ }\textbf {\bibinfo {volume} {07}},\ \bibinfo {pages} {066} (\bibinfo
  {year} {2005})},\ \Eprint {http://arxiv.org/abs/hep-th/0505160}
  {arXiv:hep-th/0505160 [hep-th]} \BibitemShut {NoStop}%
%%CITATION = HEP-TH/0505160;%%
\bibitem [{\citenamefont {Palti}(2006)}]{Palti:2006yz}%
  \BibitemOpen
  \bibfield  {author} {\bibinfo {author} {\bibfnamefont {E.}~\bibnamefont
  {Palti}},\ }\emph {\bibinfo {title} {{Aspects of moduli stabilisation in
  string and M-theory}}},\ \href@noop {} {Ph.D. thesis},\ \bibinfo  {school}
  {Sussex U.} (\bibinfo {year} {2006}),\ \Eprint
  {http://arxiv.org/abs/hep-th/0608033} {arXiv:hep-th/0608033 [hep-th]}
  \BibitemShut {NoStop}%
%%CITATION = HEP-TH/0608033;%%
\bibitem [{\citenamefont {Gauntlett}\ and\ \citenamefont
  {Varela}(2007)}]{Gauntlett:2007ma}%
  \BibitemOpen
  \bibfield  {author} {\bibinfo {author} {\bibfnamefont {J.~P.}\ \bibnamefont
  {Gauntlett}}\ and\ \bibinfo {author} {\bibfnamefont {O.}~\bibnamefont
  {Varela}},\ }\href {\doibase 10.1103/PhysRevD.76.126007} {\bibfield
  {journal} {\bibinfo  {journal} {Phys. Rev.}\ }\textbf {\bibinfo {volume}
  {D76}},\ \bibinfo {pages} {126007} (\bibinfo {year} {2007})},\ \Eprint
  {http://arxiv.org/abs/0707.2315} {arXiv:0707.2315 [hep-th]} \BibitemShut
  {NoStop}%
%%CITATION = ARXIV:0707.2315;%%
\bibitem [{\citenamefont {Micu}\ \emph {et~al.}(2007)\citenamefont {Micu},
  \citenamefont {Palti},\ and\ \citenamefont {Tasinato}}]{Micu:2007rd}%
  \BibitemOpen
  \bibfield  {author} {\bibinfo {author} {\bibfnamefont {A.}~\bibnamefont
  {Micu}}, \bibinfo {author} {\bibfnamefont {E.}~\bibnamefont {Palti}}, \ and\
  \bibinfo {author} {\bibfnamefont {G.}~\bibnamefont {Tasinato}},\ }\href
  {\doibase 10.1088/1126-6708/2007/03/104} {\bibfield  {journal} {\bibinfo
  {journal} {JHEP}\ }\textbf {\bibinfo {volume} {03}},\ \bibinfo {pages} {104}
  (\bibinfo {year} {2007})},\ \Eprint {http://arxiv.org/abs/hep-th/0701173}
  {arXiv:hep-th/0701173 [hep-th]} \BibitemShut {NoStop}%
%%CITATION = HEP-TH/0701173;%%
\bibitem [{\citenamefont {Palti}(2007)}]{Palti:2007pm}%
  \BibitemOpen
  \bibfield  {author} {\bibinfo {author} {\bibfnamefont {E.}~\bibnamefont
  {Palti}},\ }\href {\doibase 10.1088/1126-6708/2007/10/011} {\bibfield
  {journal} {\bibinfo  {journal} {JHEP}\ }\textbf {\bibinfo {volume} {10}},\
  \bibinfo {pages} {011} (\bibinfo {year} {2007})},\ \Eprint
  {http://arxiv.org/abs/0707.1595} {arXiv:0707.1595 [hep-th]} \BibitemShut
  {NoStop}%
%%CITATION = ARXIV:0707.1595;%%
\bibitem [{\citenamefont {Kachru}\ \emph {et~al.}(2003)\citenamefont {Kachru},
  \citenamefont {Kallosh}, \citenamefont {Linde},\ and\ \citenamefont
  {Trivedi}}]{Kachru:2003aw}%
  \BibitemOpen
  \bibfield  {author} {\bibinfo {author} {\bibfnamefont {S.}~\bibnamefont
  {Kachru}}, \bibinfo {author} {\bibfnamefont {R.}~\bibnamefont {Kallosh}},
  \bibinfo {author} {\bibfnamefont {A.~D.}\ \bibnamefont {Linde}}, \ and\
  \bibinfo {author} {\bibfnamefont {S.~P.}\ \bibnamefont {Trivedi}},\ }\href
  {\doibase 10.1103/PhysRevD.68.046005} {\bibfield  {journal} {\bibinfo
  {journal} {Phys. Rev.}\ }\textbf {\bibinfo {volume} {D68}},\ \bibinfo {pages}
  {046005} (\bibinfo {year} {2003})},\ \Eprint
  {http://arxiv.org/abs/hep-th/0301240} {arXiv:hep-th/0301240 [hep-th]}
  \BibitemShut {NoStop}%
%%CITATION = HEP-TH/0301240;%%
\bibitem [{\citenamefont {Balasubramanian}\ \emph {et~al.}(2005)\citenamefont
  {Balasubramanian}, \citenamefont {Berglund}, \citenamefont {Conlon},\ and\
  \citenamefont {Quevedo}}]{Balasubramanian:2005zx}%
  \BibitemOpen
  \bibfield  {author} {\bibinfo {author} {\bibfnamefont {V.}~\bibnamefont
  {Balasubramanian}}, \bibinfo {author} {\bibfnamefont {P.}~\bibnamefont
  {Berglund}}, \bibinfo {author} {\bibfnamefont {J.~P.}\ \bibnamefont
  {Conlon}}, \ and\ \bibinfo {author} {\bibfnamefont {F.}~\bibnamefont
  {Quevedo}},\ }\href {\doibase 10.1088/1126-6708/2005/03/007} {\bibfield
  {journal} {\bibinfo  {journal} {JHEP}\ }\textbf {\bibinfo {volume} {03}},\
  \bibinfo {pages} {007} (\bibinfo {year} {2005})},\ \Eprint
  {http://arxiv.org/abs/hep-th/0502058} {arXiv:hep-th/0502058 [hep-th]}
  \BibitemShut {NoStop}%
%%CITATION = HEP-TH/0502058;%%
\bibitem [{\citenamefont {Palti}\ \emph {et~al.}(2008)\citenamefont {Palti},
  \citenamefont {Tasinato},\ and\ \citenamefont {Ward}}]{Palti:2008mg}%
  \BibitemOpen
  \bibfield  {author} {\bibinfo {author} {\bibfnamefont {E.}~\bibnamefont
  {Palti}}, \bibinfo {author} {\bibfnamefont {G.}~\bibnamefont {Tasinato}}, \
  and\ \bibinfo {author} {\bibfnamefont {J.}~\bibnamefont {Ward}},\ }\href
  {\doibase 10.1088/1126-6708/2008/06/084} {\bibfield  {journal} {\bibinfo
  {journal} {JHEP}\ }\textbf {\bibinfo {volume} {06}},\ \bibinfo {pages} {084}
  (\bibinfo {year} {2008})},\ \Eprint {http://arxiv.org/abs/0804.1248}
  {arXiv:0804.1248 [hep-th]} \BibitemShut {NoStop}%
%%CITATION = ARXIV:0804.1248;%%
\bibitem [{\citenamefont {Louis}\ and\ \citenamefont
  {L{\"u}st}(2017)}]{Louis:2016tnz}%
  \BibitemOpen
  \bibfield  {author} {\bibinfo {author} {\bibfnamefont {J.}~\bibnamefont
  {Louis}}\ and\ \bibinfo {author} {\bibfnamefont {S.}~\bibnamefont
  {L{\"u}st}},\ }\href {\doibase 10.1007/JHEP02(2017)085} {\bibfield  {journal}
  {\bibinfo  {journal} {JHEP}\ }\textbf {\bibinfo {volume} {02}},\ \bibinfo
  {pages} {085} (\bibinfo {year} {2017})},\ \Eprint
  {http://arxiv.org/abs/1607.08249} {arXiv:1607.08249 [hep-th]} \BibitemShut
  {NoStop}%
%%CITATION = ARXIV:1607.08249;%%
\bibitem [{\citenamefont {L{\"u}st}\ \emph {et~al.}(2018)\citenamefont
  {L{\"u}st}, \citenamefont {Ruter},\ and\ \citenamefont
  {Louis}}]{Lust:2017aqj}%
  \BibitemOpen
  \bibfield  {author} {\bibinfo {author} {\bibfnamefont {S.}~\bibnamefont
  {L{\"u}st}}, \bibinfo {author} {\bibfnamefont {P.}~\bibnamefont {Ruter}}, \
  and\ \bibinfo {author} {\bibfnamefont {J.}~\bibnamefont {Louis}},\ }\href
  {\doibase 10.1007/JHEP03(2018)019} {\bibfield  {journal} {\bibinfo  {journal}
  {JHEP}\ }\textbf {\bibinfo {volume} {03}},\ \bibinfo {pages} {019} (\bibinfo
  {year} {2018})},\ \Eprint {http://arxiv.org/abs/1711.06180} {arXiv:1711.06180
  [hep-th]} \BibitemShut {NoStop}%
%%CITATION = ARXIV:1711.06180;%%
\bibitem [{\citenamefont {Bena}\ \emph {et~al.}(2019)\citenamefont {Bena},
  \citenamefont {Dudas}, \citenamefont {Grana},\ and\ \citenamefont
  {L{\"u}st}}]{Bena:2018fqc}%
  \BibitemOpen
  \bibfield  {author} {\bibinfo {author} {\bibfnamefont {I.}~\bibnamefont
  {Bena}}, \bibinfo {author} {\bibfnamefont {E.}~\bibnamefont {Dudas}},
  \bibinfo {author} {\bibfnamefont {M.}~\bibnamefont {Grana}}, \ and\ \bibinfo
  {author} {\bibfnamefont {S.}~\bibnamefont {L{\"u}st}},\ }\href {\doibase
  10.1002/prop.201800100} {\bibfield  {journal} {\bibinfo  {journal} {Fortsch.
  Phys.}\ }\textbf {\bibinfo {volume} {67}},\ \bibinfo {pages} {1800100}
  (\bibinfo {year} {2019})},\ \Eprint {http://arxiv.org/abs/1809.06861}
  {arXiv:1809.06861 [hep-th]} \BibitemShut {NoStop}%
%%CITATION = ARXIV:1809.06861;%%
\bibitem [{\citenamefont {Freivogel}\ \emph {et~al.}(2006)\citenamefont
  {Freivogel}, \citenamefont {Hubeny}, \citenamefont {Maloney}, \citenamefont
  {Myers}, \citenamefont {Rangamani},\ and\ \citenamefont
  {Shenker}}]{Freivogel:2005qh}%
  \BibitemOpen
  \bibfield  {author} {\bibinfo {author} {\bibfnamefont {B.}~\bibnamefont
  {Freivogel}}, \bibinfo {author} {\bibfnamefont {V.~E.}\ \bibnamefont
  {Hubeny}}, \bibinfo {author} {\bibfnamefont {A.}~\bibnamefont {Maloney}},
  \bibinfo {author} {\bibfnamefont {R.~C.}\ \bibnamefont {Myers}}, \bibinfo
  {author} {\bibfnamefont {M.}~\bibnamefont {Rangamani}}, \ and\ \bibinfo
  {author} {\bibfnamefont {S.}~\bibnamefont {Shenker}},\ }\href {\doibase
  10.1088/1126-6708/2006/03/007} {\bibfield  {journal} {\bibinfo  {journal}
  {JHEP}\ }\textbf {\bibinfo {volume} {03}},\ \bibinfo {pages} {007} (\bibinfo
  {year} {2006})},\ \Eprint {http://arxiv.org/abs/hep-th/0510046}
  {arXiv:hep-th/0510046 [hep-th]} \BibitemShut {NoStop}%
%%CITATION = HEP-TH/0510046;%%
\bibitem [{\citenamefont {Higuchi}(1987)}]{Higuchi:1986py}%
  \BibitemOpen
  \bibfield  {author} {\bibinfo {author} {\bibfnamefont {A.}~\bibnamefont
  {Higuchi}},\ }\href {\doibase 10.1016/0550-3213(87)90691-2} {\bibfield
  {journal} {\bibinfo  {journal} {Nucl. Phys.}\ }\textbf {\bibinfo {volume}
  {B282}},\ \bibinfo {pages} {397} (\bibinfo {year} {1987})}\BibitemShut
  {NoStop}%
%%CITATION = NUPHA,B282,397;%%
\bibitem [{\citenamefont {Agrawal}\ \emph {et~al.}(2018)\citenamefont
  {Agrawal}, \citenamefont {Obied}, \citenamefont {Steinhardt},\ and\
  \citenamefont {Vafa}}]{Agrawal:2018own}%
  \BibitemOpen
  \bibfield  {author} {\bibinfo {author} {\bibfnamefont {P.}~\bibnamefont
  {Agrawal}}, \bibinfo {author} {\bibfnamefont {G.}~\bibnamefont {Obied}},
  \bibinfo {author} {\bibfnamefont {P.~J.}\ \bibnamefont {Steinhardt}}, \ and\
  \bibinfo {author} {\bibfnamefont {C.}~\bibnamefont {Vafa}},\ }\href {\doibase
  10.1016/j.physletb.2018.07.040} {\bibfield  {journal} {\bibinfo  {journal}
  {Phys. Lett.}\ }\textbf {\bibinfo {volume} {B784}},\ \bibinfo {pages} {271}
  (\bibinfo {year} {2018})},\ \Eprint {http://arxiv.org/abs/1806.09718}
  {arXiv:1806.09718 [hep-th]} \BibitemShut {NoStop}%
%%CITATION = ARXIV:1806.09718;%%
\bibitem [{\citenamefont {Agrawal}\ \emph {et~al.}()\citenamefont {Agrawal},
  \citenamefont {Obied},\ and\ \citenamefont {Vafa}}]{AOV}%
  \BibitemOpen
  \bibfield  {author} {\bibinfo {author} {\bibfnamefont {P.}~\bibnamefont
  {Agrawal}}, \bibinfo {author} {\bibfnamefont {G.}~\bibnamefont {Obied}}, \
  and\ \bibinfo {author} {\bibfnamefont {C.}~\bibnamefont {Vafa}},\ }\href@noop
  {} {\bibinfo  {journal} {To Appear}\ }\BibitemShut {NoStop}%
\bibitem [{\citenamefont {Alday}\ and\ \citenamefont
  {Perlmutter}(2019)}]{Alday:2019qrf}%
  \BibitemOpen
\bibfield  {journal} {  }\bibfield  {author} {\bibinfo {author} {\bibfnamefont
  {L.~F.}\ \bibnamefont {Alday}}\ and\ \bibinfo {author} {\bibfnamefont
  {E.}~\bibnamefont {Perlmutter}},\ }\href@noop {} {\  (\bibinfo {year}
  {2019})},\ \Eprint {http://arxiv.org/abs/1906.01477} {arXiv:1906.01477
  [hep-th]} \BibitemShut {NoStop}%
%%CITATION = ARXIV:1906.01477;%%
\end{thebibliography}%

\end{document}